\documentclass{amsart}

\usepackage{amsmath,amssymb,amsfonts,amscd,amsthm,amsopn,epsfig}  
\usepackage{amsthm}
\usepackage{amssymb}
\usepackage{latexsym,verbatim}
\usepackage{mathrsfs}
\usepackage{enumerate}

\usepackage{tikz}\usetikzlibrary{backgrounds,matrix,arrows,shapes,decorations.markings}
\newcommand{\btp}{\begin{tikzpicture}[baseline=-5pt,scale=0.25,line width=0.7pt]}
\newcommand{\etp}{\end{tikzpicture}}

\pgfdeclarelayer{background}
\pgfdeclarelayer{foreground}
\pgfsetlayers{background,main,foreground}

\usepackage{hyperref}
\hypersetup{
    citecolor=blue,        
}

\usepackage{color}

\newtheorem{rmk}{Remark}[section]

\numberwithin{equation}{section}

\def\r{\rho}

\def\V={{{\bf\rm{V}}}}

\def\beq{\begin{equation}}
\def\eeq{\end{equation}}
\def\bea{\begin{eqnarray}}
\def\eea{\end{eqnarray}}
\def\ba{\begin{array}}
\def\ea{\end{array}}

\def\lt{\left}
\def\rt{\right}

\def\beq{\begin{equation}}
\def\eeq{\end{equation}}
\def\ben{\begin{eqnarray}}
\def\een{\end{eqnarray}}
\def\ba{\begin{array}}
\def\ea{\end{array}}

\def\lt{\left}
\def\rt{\right}

\newcommand\cH{{\mathcal H}}

\newcommand\CC{\mathbb C}

\newcommand{\so}{\scriptscriptstyle \rm I}
\newcommand{\st}{\scriptscriptstyle \rm I\hspace{-1pt}I}

\begin{document}

\title{Modified algebraic Bethe ansatz for XXZ chain on the segment\\  - II - general cases}
\author{S.~Belliard}
\address{Laboratoire de Physique Th\'eorique
et Mod\'elisation (CNRS UMR 8089), Universit\'e de Cergy-Pontoise, F-95302 Cergy-Pontoise, France }
\email{samuel.belliard@u-cergy.fr}
\author{R.A.~Pimenta}
\address{Universidade Federal de S\~ao Carlos, Departamento de F\'{\i}sica\\
Caixa Postal 676, CEP 13569-905, S\~ao Carlos, Brasil}
\email{pimenta@df.ufscar.br}

\begin{abstract}

The spectral problem of the Heisenberg XXZ spin-$\frac{1}{2}$ chain on the
segment is investigated within a modified algebraic Bethe ansatz framework. We consider
in this work the most general boundaries allowed by integrability.
The eigenvalues and the eigenvectors
are obtained. They are characterised by a set of Bethe roots with cardinality
equal to $N$, the
length of the chain, and which satisfies a set of Bethe equations with an additional term.

\end{abstract}

\maketitle

\vskip -0.2cm

{\small  MSC: 82B23; 81R12}

{{\small  {\it \bf Keywords}: algebraic Bethe ansatz; integrable spin chain; boundary conditions}}

\section{Introduction}

In this work, we study the spectral problem of the Heisenberg XXZ
spin-$\frac{1}{2}$ chain on the segment,
namely
\ben\label{H}
&&H=
\epsilon\, \sigma^{z}_1 +  \kappa^-\,\sigma^{-}_1  +  \kappa^+\,\sigma^{+}_1   +
\sum_{k=1}^{N-1}\Big(
\sigma^{x}_{k}\otimes \sigma^{x}_{k+1}+\sigma^{y}_{k}\otimes \sigma^{y}_{k+1} +
\Delta\sigma^{z}_{k}\otimes \sigma^{z}_{k+1}\Big)  +
\nu\,  \sigma^{z}_N +\tau^-\,  \sigma^{-}_N+\tau^+\,  \sigma^{+}_N,
\een
where $N$ is the length of the chain and $\sigma_i^{\pm,x,y,z}$
are the standard Pauli matrices,
which act non-trivially on the site $i$
of the quantum space $\cH= \otimes_{i=1}^N \CC^2$.
The theory is characterised by the anisotropy parameter $\Delta=\frac{q+q^{-1}}{2}$,
where $q$ is generic,
and by left  $\{ \epsilon,  \kappa^\pm\}$ and right 
$\{ \nu,  \tau^\pm\}$ boundary couplings that we consider generic.

\vspace{0.2cm}

The possibility of exactly solving the Hamiltonian (\ref{H}) is due to the fact
that it can be embedded into the quantum inverse scattering framework \cite{Skl88}. Indeed,
the spin-$\frac{1}{2}$ chain on the segment can be obtained from the logarithmic derivative
of the double-row transfer matrix of the lattice six vertex model with reflected ends.
The commutativity of the double-row transfer matrix, which allows one to diagonalise the Hamiltonian and
the transfer matrix in a common basis, follows from the so-called reflection equations \cite{Che84},
in addition to the Yang-Baxter equation.

\vspace{0.2cm}

The complete characterisation of the spectrum of the double-row transfer
matrix is, however, a challenging problem, which has been investigated
by many authors, either
from a Bethe ansatz (BA) point of view
\cite{TQ,TQ2,gauge,GP04,GNPR05,YanZ07,CRS1,CRS2,BCR12,PL13,PL14,CYSW3,KMN14}
or from alternative approaches
\cite{BK,Gal08,niccoli2,BB,FKN,LP14}, see for instance \cite{Bel14} for more historical details.
The main difficulty arises from the breaking of the $U(1)-$symmetry, here induced by
general boundary configurations allowed by the reflection equations.
Let us also mention that, besides its intrinsic interest, the solution of this problem also find
applications in other areas such as out-of-equilibrium statistical
physics, high energy physics, condensed matter, mathematical physics, among others.

\vspace{0.2cm}

In a previous paper \cite{Bel14}, one of the authors considered the Hamiltonian (\ref{H})
with triangular boundaries, {\it i.e.}, $\kappa^+=\tau^+=0$ or
$\kappa^-=\tau^+=0$,
from the so-called modified algebraic Bethe ansatz (MABA) approach. This method
was first implemented in the case of the 
XXX  spin-$\frac{1}{2}$ chain on the segment with the most general boundary conditions
\cite{BC13} and recently applied to the totally asymmetric exclusion process \cite{Cra14}.
By means of this method, the algebraic Bethe ansatz framework \cite{SFT,Skl88} can now be used
for quantum integrable models with finite number of degree of freedom
and which are not invariant by the $U(1)-$symmetry.

\vspace{0.2cm}

Here, we consider the Heisenberg XXZ  spin-$\frac{1}{2}$ chain on the segment
with generic boundary parameters. We obtain constructively the Bethe vectors, the eigenvalues
(or equivalently the functional Baxter T-Q equation)
and the Bethe equations. We recover the Baxter T-Q
equation with the new additional term discovered in \cite{CYSW3, KMN14}.
The new results are the construction of the associated Bethe vectors
which characterise the
eigenstates of the Hamiltonian (\ref{H}) as well as the off-shell action of the transfer matrix on it.

\vspace{0.2cm}

We also consider the limiting cases 
with left general
and right upper triangular boundaries, {\it i.e.} $\tau^+=0$, and the case where the boundary parameters satisfy certain
constraints \cite{TQ,TQ2,gaugeKF}, which can be obtained by requesting the vanishing
of extra unwanted terms in the off-shell action of the transfer matrix on the Bethe vectors.

\vspace{0.2cm}

This paper is organised as follows. 
In the section \ref{RKRKform}, the basic properties
of the quantum group $U_q(\widehat {sl_2})$ and
of its coideal sub-algebra are reminded.
Next, in section \ref{S:Gauge}, we recall the dynamical gauge transformations
which allow the construction of the dynamical operators and of the dynamical transfer matrix.
In section \ref{S:Rep}, the representation theory, necessary for
the implementation of the MABA, is considered.
Then, in section \ref{S:MABA}, we develop the modified algebraic Bethe ansatz and in section \ref{S:Lim} we give some limit cases of our result.
Finally, we conclude in the section \ref{S:Conc}.



\section{XXZ chain on the segment in the reflection algebra formalism  \label{RKRKform} }

Quantum integrable models can be constructed from quantum groups and their coideal sub-algebras,
the so-called reflection algebras. For the XXZ spin chain on the segment, one consider the quantum group
$U_q(\widehat{sl_2})$ and its reflection algebras \cite{Che84,Skl88}.
Here we give only the needed relations, more details can be found for instance in \cite{Bel14}.
The fundamental object in this context is the so-called $R-$matrix, which acts on $V_a\otimes V_b$\footnote{In this work, we only use two-dimensional complex vector spaces, {\it i.e.}, $V=\CC^2$.}, and it is given by,
\ben\label{R}
\qquad R_{ab}(u)=\lt(\begin{array}{cccc}b(qu)&0&0&0\\0&b(u)&1&0\\
0&1&b(u)&0\\0&0&0&b(qu)\end{array}\rt),\quad
b(u)=\frac{u-u^{-1}}{q-q^{-1}},
\een
and which satisfies the quantum Yang-Baxter equation that acts on $V_a\otimes V_b \otimes V_c$,
\ben\label{YB}
R_{ab}(u_a/u_b)R_{ac}(u_a/u_c)R_{bc}(u_b/u_c)
=R_{bc}(u_b/u_c)R_{ac}(u_a/u_c)R_{ab}(u_a/u_b).
\een 
In order to consider spin chains on the segment,
we also need the so-called $K^--$matrix and its dual, the $K^+-$matrix, which are
given by
\footnote{For convenience, we use squared
off-diagonal parameters of the $K-$matrices when compared with
the notation in \cite{Bel14}.},
\ben\label{Km}
&&K^-(u)=\lt(\begin{array}{cc}k^-(u)&\tau^2 \,c(u)\\
\tilde \tau^2 \,c(u)&k^-(u^{-1})\end{array}\rt),
\quad k^-(u) =\nu_-u+\nu_+u^{-1},\quad  c(u)=u^2-u^{-2},\\
\label{Kp}
&&K^+(u)=
\lt(\begin{array}{cc}
k^+(qu)&\tilde \kappa^2 \,c(qu)\\
\kappa^2 \,c(qu)&k^+(q^{-1}u^{-1})
\end{array}\rt),\quad k^+(u) =\epsilon_+u+\epsilon_-u^{-1},
\een
where $\{\epsilon_\pm,\kappa, \tilde \kappa \}$
and $\{\nu_\pm,\tau,\tilde \tau \}$ are generic parameters.
They are the most general solutions \cite{dVGR1} of
the reflection equation \cite{Che84} and of the dual reflection equation \cite{Skl88}, that acts on $V_a\otimes V_b$,
\ben \label{RE}
 &&R_{ab}(u_1/u_2)K^-_a(u_1)R_{ab}(u_1u_2)K^-_b(u_2)=
K^-_b(u_2)R_{ab}(u_1u_2)K^-_a(u_1)R_{ab}(u_1/u_2),\\
\label{DRE}
&&R_{ab}(u_2/u_1)K^+_a(u_1)R_{ab}(q^{-2}u^{-1}_1u^{-1}_2)K^+_b(u_2)=
K^+_b(u_2)R_{ab}(q^{-2}u^{-1}_1u^{-1}_2)K^+_a(u_1)R_{ab}(u_2/u_1).
\een 

The $R-$matrix (\ref{R}) and the $K^--$matrix (\ref{Km}) allow the construction of the 
so-called double-row monodromy matrix, that acts on $V_a\otimes \cH$, given by,
\ben\label{K}
K_a(u)&=&R_{a1}(u/v_1)\dots R_{aN}(u/v_N)K_a^-(u)R_{aN}(uv_N)\dots R_{a1}(uv_1),\\
&=&\left(\begin{array}{cc}
       \mathscr{A}(u) & \mathscr{B}(u)\\
       \mathscr{C}(u) & \mathscr{D}(u)+\frac{1}{b(qu^2)}\mathscr{A}(u)
      \end{array}
\right)_a\label{KO},
\een 
with operator entries in the auxiliary space and denoted by
$\{ \mathscr{A}(u) , \mathscr{B}(u), \mathscr{C}(u) , \mathscr{D}(u)\}$.
Each of these operators act on the quantum space $\cH=V_1\otimes \dots \otimes V_N$. This
representation is useful in the construction of the eigenstates
by means of the $\mathscr{B}(u)$ operator.
The parameters $v_i$ in (\ref{K}) are called inhomogeneities.
\begin{rmk}
It is also possible to use another family of operators,
$\{ \mathscr{\hat A}(u) , \mathscr{B}(u), \mathscr{C}(u) , \mathscr{\hat D}(u)\}$,
which allow to construct the Bethe vectors of the transfer matrix
from the $\mathscr{C}(u)$ operator,
\ben\label{KOt}
K_a(u)&=&\left(\begin{array}{cc}
       \mathscr{\hat A}(u)+\frac{1}{b(qu^{2})}\mathscr{\hat D}(u) & \mathscr{B}(u)\\
       \mathscr{C}(u) & \mathscr{\hat D}(u)
      \end{array}
\right)_a.
\een 
\end{rmk}
Bringing together the dual $K^+-$matrix (\ref{Kp}) and the double-row monodromy matrix
(\ref{KO}), one can construct the double-row transfer matrix, 
\ben\label{tr}
\qquad t(u)=tr_a(K^+_a(u)K_a(u))=
\phi(u)k^+(u)\mathscr{A}(u) +
k^+(q^{-1}u^{-1})\mathscr{D}(u)+c(qu)
\Big(\kappa^2 \,\mathscr{B}(u)+\tilde\kappa^2 \,\mathscr{C}(u)\Big),
\een
where
\ben\label{phi}
\phi(u)= \frac{b(q^2u^2)}{b(qu^2)},
\een
which is the operator whose spectrum will be investigated.
\begin{rmk} In a similar way, we can use the operators
$\{ \mathscr{\hat A}(u), \mathscr{\hat D}(u)\}$ to write the transfer matrix as,
\ben\label{trt}
t(u)= k^+(q u)\mathscr{\hat A}(u) +\phi(u)k^+(u^{-1})\mathscr{\hat D}(u)+
c(qu)\Big(\kappa^2 \,\mathscr{B}(u)+\tilde\kappa^2 \,\mathscr{C}(u)\Big).
\een
\end{rmk}
The transfer matrix, since the Yang-Baxter and reflection
equation are satisfied, constitutes a family of commuting operators, if evaluated at
different spectral parameters \cite{Skl88},
{\it i.e.} $[t(u),t(v)]=0$. For this reason,
$t(u)$ can be regarded as a generating function of the conserved charges of the model.
In this context, the Hamiltonian (\ref{H}) is one of these charges
and it is related to the (homogeneous)
double-row transfer matrix
by means of the relation,
\ben\label{Htotr}
H=\frac{q-q^{-1}}{2} \frac{d}{du}\ln(t(u))
\Big|_{u=1, v_i=1}-
\left(
N~\frac{q+q^{-1}}{2}+\frac{(q-q^{-1})^2}{2(q+q^{-1})}
\right).
\een

In terms of the boundary parameters of the $K-$matrices (\ref{Km},\ref{Kp}),
the couplings of the Hamiltonian (\ref{H}) are expressed as,
\ben\label{parKH1}
&&\epsilon=\frac{(q-q^{-1})}{2}\frac{(\epsilon_+ - \epsilon_-)}{(\epsilon_+ + \epsilon_-)}
,\quad
  \kappa^-=\frac{2(q-q^{-1})}{(\epsilon_+ + \epsilon_-)}\kappa^2,
  \quad   \kappa^+=\frac{2(q-q^{-1})}{(\epsilon_+ + \epsilon_-)}\tilde \kappa^2,\\
  \label{parKH2}
&& \nu=  \frac{(q-q^{-1})}{2}\frac{(\nu_- - \nu_+)}{(\nu_+ +\nu_-)},\quad
 \tau^-= \frac{2(q-q^{-1})}{(\nu_+ + \nu_-)}\tilde \tau^2,\quad
 \tau^+= \frac{2(q-q^{-1})}{(\nu_+ + \nu_-)}\tau^2.
 \een
 
It is convenient to introduce a new parametrisation for the boundary parameters,
namely
\ben\label{NpKm}
&& \nu_-=i\tilde\tau\tau\big(\mu/\tilde \mu+\tilde \mu/\mu\big),
 \quad \nu_+=i\tilde\tau\tau\big(\mu\tilde\mu+1/( \mu\tilde \mu)\big),
\\ \label{NpKp} &&\epsilon_-=i\tilde\kappa\kappa\big(\xi/\tilde \xi+\tilde\xi/ \xi\big),
\quad \epsilon_+=i\tilde\kappa\kappa\big(\xi \tilde \xi+1/(\tilde\xi \xi)\big).
\een
The advantage of this parametrisation is that it
brings the $q-$determinant \cite{Skl88}
to a factorized structure, \textit{i.e.},
\bea
&&{\rm Det}_q\{K^-(u)\}=
tr_{12}
\Big(P^{-}_{12}K^-_1(u)R_{12}(qu^2)K^-_2(qu)\Big)
=
b(u^2)\tilde k^-(qu)\tilde k^-(q^{-1}u^{-1}),
\\
&&{\rm Det}_q\{K^+(u)\}=
tr_{12}
\Big(P^{-}_{12}K^+_2(qu)R_{12}(q^{-3}u^{-2})K^+_1(u)\Big)
=
b(q^{-4}u^{-2})\tilde k^+(qu)\tilde k^+(q^{-1}u^{-1}),
\eea
where we introduce useful boundary functions given by,
\bea
\tilde k^-(u)=
i \tilde\tau\tau ( \mu u +\mu^{-1} u^{-1})(\tilde\mu^{-1}u + \tilde\mu u^{-1}), \quad
\tilde k^+(u)=
i \tilde\kappa\kappa   ( \tilde \xi u +\tilde \xi^{-1} u^{-1})(\xi^{-1}u + \xi u^{-1}).
\eea

\begin{rmk}\label{Z2par}
The parametrisations (\ref{NpKm},\ref{NpKp}) are respectively invariant
by a $Z^2\times Z^2$ symmetry, namely
\begin{itemize}
\item For (\ref{NpKm}) by the transformation  $\mu \to \tilde \mu$ and $\tilde \mu \to \mu$ or  $ \tilde \mu \to 1/ \tilde \mu $ and $\mu \to 1/ \mu $. 
\item  For (\ref{NpKp}) by the transformation  $\xi \to \tilde \xi$ and  $\tilde \xi \to \xi$  or    $ \tilde \xi \to 1/ \tilde \xi$ and $\xi \to 1/ \xi$.
\end{itemize}
\end{rmk}



\section{Dynamical gauge transformation of the transfer matrix \label{S:Gauge}}

In the case of the XXX spin-$\frac{1}{2}$ chain on the segment, due to the $SU(2)$ invariance,
similarity transformations can be used
to map the original problem with general boundaries to another one with a left general
and a right diagonal boundaries or to a
left lower triangular and right upper
triangular boundaries. Both cases can be considered within the MABA framework \cite{BC13,Bel14}.

\vspace{0.2cm}

For the XXZ case, the $SU(2)$ invariance is broken
by the presence of the anisotropy parameter
and one has to consider a more intricate transformation \cite{gauge}.
We recall that this transformation was introduced by Baxter in \cite{baxter}
to study the XYZ spin chain on the circle (see also \cite{FT79}) and
then used in the context of the XXZ spin chain on the segment in \cite{gauge}. 
As we shall demonstrate, the local gauge transformation allows one to bring
the dynamical double-row transfer matrix to a dynamical lower-upper (or upper-lower) structure.

\vspace{0.2cm}

In this section, after defining the gauge transformation, we introduce the dynamical operators
and the dynamical transfer matrix. Next, the important multiple commutation relations
for the MABA are also derived.

\subsection{Gauge vectors} Following \cite{gauge}, we introduce the covariant vectors
and the contravariant vectors,
\bea\label{coVec}
&& X(u,m)=\left(\begin{array}{c}
     \alpha q^{-m} u^{-1} \\
       1
      \end{array}
\right),\quad
Y(u,m)=\left(\begin{array}{c}
     \beta q^{m} u^{-1} \\
       1
      \end{array}
\right),\\\label{conVec}
&& \tilde X(u,m)=\frac{q u}{\gamma_{m-1}}\left(\begin{array}{cc}
-1,& \alpha q^{-m} u^{-1}
\end{array}
\right),\quad
\tilde Y(u,m)=\frac{q u}{\gamma_{m+1}}\left(\begin{array}{cc}
1 ,&
      - \beta q^{m} u^{-1} 
      \end{array}
\right),
\eea
where $\alpha$ and $\beta$ are arbitrary complex parameters and
$m$ is an integer which characterises the dynamical algebra produced by the
gauge transformation (see \textit{e.g.} \cite{YanZ07,YCHHSZ11,gaugeKF} for more details).
They satisfy the scalar products,
\bea\label{Scal-prod}
&&\tilde X(u,m)X(u,m)=\tilde Y(u,m)Y(u,m)=0, \quad \tilde X(u,m+1)Y(u,m-1)=\tilde Y(u,m-1)X(u,m+1)=1
\eea
as well as the closure relation,
\bea\label{close}
&& Y(u,m-1)\tilde X(u,m+1)+X(u,m+1)\tilde Y(u,m-1)= \left(\begin{array}{cc}
       1 & 0\\
      0 & 1      \end{array}\right).
\eea
Let us introduce the function $\gamma(u,m)=\alpha q^{-m} u-\beta q^{m} u^{-1}$
and denote $ \gamma(1,m)=\gamma_m$.
The covariant vectors then satisfy the following intertwining relations with the $R-$matrix (\ref{R}),
\ben
&&R_{12}(u/v)X_1(u,m+1) \otimes X_2(v,m)=b(qu/v)X_1(u,m)  \otimes X_2(v,m+1),\label{RXY}\\
&&R_{12}(u/v)Y_1(u,m)  \otimes Y_2(v,m+1)=b(qu/v)Y_1(u,m+1)  \otimes Y_2(v,m),\nonumber\\
&&R_{12}(u/v)X_1(u,m+1)  \otimes Y_2(v,m)=\frac{b(u/v)\gamma_{m}}{\gamma_{m+1}}X_1(u,m+2)  \otimes Y_2(v,m+1)\nonumber\\
&&\qquad\qquad\qquad\qquad\qquad\qquad\qquad\qquad\qquad+\frac{\gamma(v/u,m+1)}{\gamma_{m+1}}Y_1(u,m)  \otimes X_2(v,m+1),\nonumber\\
&&R_{12}(u/v)Y_1(u,m)  \otimes X_2(v,m+1)=\frac{b(u/v)\gamma_{m+1}}{\gamma_{m}}Y_1(u,m-1)  \otimes X_2(v,m)\nonumber\\
&&\qquad\qquad\qquad\qquad\qquad\qquad\qquad\qquad\qquad+\frac{\gamma(u/v,m)}{\gamma_{m}}X_1(u,m+1)  \otimes Y_2(v,m)\nonumber
\een
while the contravariant relations are given by,
\ben
&&\tilde X_1(u,m+1)  \otimes \tilde X_2(v,m)R_{12}(u/v)=b(qu/v)\tilde X_1(u,m)  \otimes \tilde X_2(v,m+1),\label{RtXtY}\\
&&\tilde Y_1(u,m)  \otimes \tilde Y_2(v,m+1)R_{12}(u/v)=b(qu/v)\tilde Y_1(u,m+1)  \otimes \tilde Y_2(v,m),\nonumber\\
&&\tilde X_1(u,m+1) \otimes \tilde Y_2(v,m-2)R_{12}(u/v)=
\frac{b(u/v)\gamma_{m+1}}{\gamma_{m}}\tilde X_1(u,m+2) \otimes\tilde Y_2(v,m-1)\nonumber\\
&&\qquad\qquad\qquad\qquad\qquad\qquad\qquad\qquad\qquad +
\frac{\gamma(v/u,m)}{\gamma_{m}}\tilde Y_1(u,m-2) \otimes \tilde X_2(v,m+1),\nonumber\\
&&\tilde Y_1(u,m-1) \otimes \tilde X_2(v,m+2)R_{12}(u/v)=\frac{b(u/v)\gamma_{m-1}}{\gamma_{m}}\tilde Y_1(u,m-2) \otimes \tilde X_2(v,m+1)\nonumber\\
&&\qquad\qquad\qquad\qquad\qquad\qquad\qquad\qquad\qquad+\frac{\gamma(u/v,m)}{\gamma_{m}}\tilde X_1(u,m+2) \otimes \tilde Y_2(v,m-1)\nonumber.
\een



\subsection{Dynamical  operators and transfer matrices}Following \cite{gauge},
we introduce the dynamical operators
\bea\label{odyn}
&&\mathscr{C}(u,m) =\tilde X(u,m)K(u)X(u^{-1},m),\quad
\mathscr{B}(u,m)= \tilde Y(u,m)K(u)Y(u^{-1},m),\\
&&\mathscr{A}(u,m)=\tilde Y(u,m-2)K(u)X(u^{-1},m), \quad
\mathscr{\hat D}(u,m)=\tilde X(u,m+2)K(u)Y(u^{-1},m),
\eea
and  
\ben
&&\mathscr{\hat A}(u,m)=
\frac{\gamma_{m-1}}{\gamma_m}\mathscr{A}(u,m)-\frac{\gamma(u^{2},m-1)}{b(qu^2)\gamma_m} \mathscr{\hat D}(u,m),\\
&&
\mathscr{D}(u,m)=
\frac{\gamma_{m+1}}{\gamma_m}\mathscr{\hat D}(u,m)-
\frac{\gamma(u^{-2},m+1)}{b(qu^2)\gamma_m} \mathscr{A}(u,m),
\een
which satisfy the commutation relations given in the appendix (\ref{App:Func}).
The family of dynamical operators
$\{\mathscr{A}(u,m),\mathscr{D}(u,m)\}$ is used to construct the Bethe vectors from the
$\mathscr{B}(u,m)$ operator. In this case, which we will consider in more details,
the transfer matrix can be decomposed in the following way,
\ben\label{t}
t(u)=t_d(u,m)+
u^{-1}c( q u)\Big( \zeta_{m}\mathscr{B}(u,m)-\tilde \zeta_m \mathscr{C}(u,m)- \delta_mt_{ps}(u,m)\Big)
\een
where
\ben
&& \zeta_m =\frac{\kappa^2}{\gamma_m}
\Big( \alpha q^{-m-1} +i\frac{\tilde\kappa\xi}{ \kappa\tilde \xi }\Big)
\Big( \alpha q^{-m-1} +i \frac{\tilde\kappa\tilde\xi}{\kappa \xi }\Big),
\quad \tilde \zeta_m =\frac{\kappa^2}{\gamma_m}
\Big( \beta q^{m-1} +i  \frac{\tilde \kappa \xi}{ \kappa \tilde \xi }\Big)
\Big( \beta q^{m-1} +i \frac{ \tilde\kappa \tilde\xi}{  \kappa \xi }\Big),\\
&& \qquad\qquad\qquad\qquad
\delta_m=
\frac{\kappa^2}{\gamma_{m+1}} \Big(\alpha q^{-m-1}
+i \frac{\tilde \kappa  \xi}{\kappa \tilde \xi }\Big)
\Big(\beta q^{m+1}+i \frac{\tilde\kappa\tilde \xi}{ \kappa\xi } \Big).
\een
The term $t_d(u,m)$, which is called the diagonal transfer matrix, is given by,
\ben\label{td}
\qquad\quad t_d(u,m)=\tilde a(u)\mathscr{A}(u,m)+\tilde d(u)\mathscr{D}(u,m),\quad 
 \tilde a (u)=u^{-1}\phi(u)\tilde k^+(u), \quad \tilde d (u)=u^{-1}\tilde k^+(q^{-1}u^{-1}),
\een
and corresponds to the only remaining contribution when there
exist constraints between left and right boundary parameters \cite{gauge}.
The term $t_{ps}(u,m)$ is defined as,
\ben\label{tps}
t_{ps}(u,m)=\phi(q^{-1}u^{-1})  \mathscr{A}(u,m)-\mathscr{D}(u,m).
\een
Although it also involves the dynamical operators $\mathscr{A}(u,m)$ and $\mathscr{D}(u,m)$,
this term disappears in the constrained case, however it has to be considered in the generic one.

\begin{rmk}
In an analogous way, we can express the transfer matrix
in terms of the operators $\{\mathscr{\hat A}(u,m),\mathscr{\hat D}(u,m)\}$. This
is convenient for the construction of the Bethe vectors from the $\mathscr{C}(u,m)$ operator.
We have,
\ben\label{th}
t(u)=\hat t_d(u,m)+u^{-1}c( q u)\Big( \zeta_{m}\mathscr{B}(u,m)-\tilde \zeta_m \mathscr{C}(u,m)+
\delta_{m-2} \hat t_{ps}(u,m)\Big)
\een
where
\ben\label{thd}
\hat t_d(u,m)=\hat a(u) \mathscr{\hat A}(u,m)+\hat d(u) \mathscr{\hat D}(u,m),\quad
\hat a (u)=u^{-1}\tilde k^+(qu), \quad
\hat d (u)=u^{-1}\phi(u)\tilde k^+(u^{-1})
\een
and
\ben\label{thps}
\hat t_{ps}(u,m)=- \mathscr{\hat A}(u,m)+\phi(q^{-1}u^{-1})   \mathscr{\hat D}(u,m).
\een
\end{rmk}

In the following subsection, we determine the action of the dynamical
transfer matrices on an ordered product of dynamical $\mathscr{B}$ (or $\mathscr{C}$) operators.
This is an fundamental step in the execution of the ABA.
In this subsection and the remaining of this paper, the set of $M$ variables $\{u_1,u_2,\dots,u_M\}$
will be shortly notated by $\bar u $ with $\# \bar u =M$.
Another important set is $\bar u_i =\{u_1,u_2,\dots,u_{i-1},u_{i+1},\dots,u_M\}$,
where the element $u_i$ is removed.
On the other hand, if an element $u$ is added to this last set we denote it by $\{u,\bar u_i\}$.
All the needed functions are given in appendix \ref{App:Func}.
In addition, for products of functions regarding to the set $\bar u$ we use a shorthand notation; for example, for the products of the function $f$, we have
\ben 
f(u,\bar u)=
\prod_{i=1}^Mf(u,u_i).\nonumber
\een
Similarly for the set $\bar u_i$ we denote 
\ben 
f(u_i,\bar u_i)=
\prod_{j=1, j \neq i}^Mf(u_i,u_j).\nonumber
\een

\subsection{Multiple commutation relations and ABA framework} Let us introduce a string of dynamical $\mathscr{B}$ operators,
with length $M$, namely
\ben\label{SB}
B(\bar u,m,M)=\mathscr{B}(u_1,m-2)\dots \mathscr{B}(u_M,m-2M).
\een
Using the dynamical commutation relations (\ref{comBdBd},\ref{comAdBd},\ref{comDdBd}),
we can show that the action of the diagonal dynamical operators
$\{\mathscr{A}(u,m),\mathscr{ D}(u,m)\}$ on this string is given by,
\ben\label{AonSB}
&&\mathscr{A}(u,m)B(\bar u,m,M)=
f(u,\bar u)B(\bar u,m,M)\mathscr{A}(u,m-2M)\\
&&\qquad\qquad+
\sum_{i=1}^M g(u,u_i,m-2)f(u_i,\bar u_i)B(\{u,\bar u_i\},m,M)\mathscr{A}(u_i,m-2M)
\nonumber
\\
&&\qquad\qquad+
\sum_{i=1}^M w(u,u_i,m-2)h(u_i,\bar u_i)B(\{u,\bar u_i\},m,M)\mathscr{D}(u_i,m-2M)
\nonumber\
\een
and
\ben\label{DonSB}
&&\mathscr{D}(u,m)B(\bar u,m,M)=
h(u,\bar u)B(\bar u,m,M)\mathscr{D}(u,m-2M)\\\nonumber
&&\qquad\qquad+\sum_{i=1}^M k(u,u_i,m-2)h(u_i,\bar u_i)B(\{u,\bar u_i\},m,M)\mathscr{D}(u_i,m-2M)\\
&&\qquad\qquad+\sum_{i=1}^M n(u,u_i,m-2)f(u_i,\bar u_i)B(\{u,\bar u_i\},m,M)\mathscr{A}(u_i,m-2M).\nonumber
\een
Then, taking into account the following functional relations among coefficients
of the commutation relations (see appendix (\ref{App:Func}) for explicit formulas),
\ben
&&\tilde a(u) g(u,v,m)+\tilde d(u) n(u,v,m)=
\tilde F(u,v)\phi(q^{-1}v^{-1})\tilde a(v)+\chi_{m+2}u^{-1} c(q u) \phi(q^{-1}v^{-1}),\\
&&\tilde a(u) w(u,v,m)+\tilde d(u) k(u,v,m)=
-\tilde F(u,v)\phi(v)\tilde d(v)-\chi_{m+2}u^{-1} c(q u),\\
&&\phi(q^{-1}u^{-1})   g(u,v,m)- n(u,v,m)=
\phi(q^{-1}v^{-1})\Big(G(u,v)b(v^2)+\r_{m+2}  \Big),\\
 &&\phi(q^{-1}u^{-1})  w(u,v,m)-k(u,v,m)=
 -\Big(G(u,v)b(q^{-2}v^{-2})+\r_{m+2} \Big),
\een
where
\ben
\chi_m=i\tilde\kappa\kappa(q-q^{-1})\frac{\gamma(\tilde\xi/\xi,m)}{\gamma_{m-1}}, \quad
\r_m=(q-q^{-1})\frac{q^{-m}\alpha+q^{m}\beta}{\gamma_{m-1}},
\een 
we find that the action of the dynamical diagonal transfer matrix (\ref{td})
on (\ref{SB}) is written as,
\ben\label{Atd}
&&t_d(u,m)B(\bar u,m,M)=B(\bar u,m,M)
\Big(
f(u,\bar u)\tilde a(u)\mathscr{A}(u,m-2M)+h(u,\bar u)\tilde d(u)\mathscr{D}(u,m-2M)
\Big)\\
&&\qquad\qquad +
\sum_{i=1}^M \tilde F(u,u_i)B(\{u,\bar u_i\},m,M)\big(\phi(q^{-1}u_i^{-1})f(u_i,\bar u_i)\tilde a(u_i)\mathscr{A}(u_i,m-2M)\nonumber\\
&&\qquad\qquad \qquad \qquad \qquad \qquad \qquad \qquad \qquad \qquad  -\phi(u_i)h(u_i,\bar u_i)\tilde d(u_i)\mathscr{D}(u_i,m-2M)\big)\nonumber\\
&&\qquad\qquad\quad +\chi_mu^{-1} c(q u)\sum_{i=1}^MB(\{u,\bar u_i\},m,M)\big(\phi(q^{-1}u_i^{-1})f(u_i,\bar u_i)\mathscr{A}(u_i,m-2M)\nonumber\\
&&\qquad\qquad \qquad \qquad \qquad \qquad \qquad \qquad \qquad \qquad  -h(u_i,\bar u_i)\mathscr{D}(u_i,m-2M)\big)\nonumber
\een
while for the action of (\ref{tps}) on (\ref{SB}) we have,
\ben\label{Atps}
&&t_{ps}(u,m)B(\bar u,m,M)=B(\bar u,m,M)\Big(\phi(q^{-1}u^{-1})f(u,\bar u)\mathscr{A}(u,m-2M)-h(u,\bar u)\mathscr{D}(u,m-2M)\Big)\\
&&\qquad\qquad +\sum_{i=1}^M G(u,u_i)B(\{u,\bar u_i\},m,M)\big(b(u_i^2)\phi(q^{-1}u_i^{-1})f(u_i,\bar u_i)\mathscr{A}(u_i,m-2M)\nonumber\\
&&\qquad\qquad \qquad \qquad \qquad \qquad \qquad \qquad \qquad \qquad  -b(q^{-2}u_i^{-2})h(u_i,\bar u_i)\mathscr{D}(u_i,m-2M)\big)\nonumber\\
&&\qquad\qquad+\r_m\sum_{i=1}^MB(\{u,\bar u_i\},m,M)\big(\phi(q^{-1}u_i^{-1})f(u_i,\bar u_i)\mathscr{A}(u_i,m-2M)\nonumber\\
&&\qquad\qquad \qquad \qquad \qquad \qquad \qquad \qquad \qquad \qquad  -h(u_i,\bar u_i)\mathscr{D}(u_i,m-2M)\big).\nonumber
\een

\begin{rmk}
Similarly, we can introduce the string of dynamical $\mathscr{C}$ operators
\ben\label{SC}
C(\bar u,m,M)=\mathscr{C}(u_1,m+2)\dots \mathscr{C}(u_M,m+2M),
\een
and, from the dynamical commutation relations (\ref{comCdCd},\ref{comhAdCd},\ref{comhDdCd}) 
and the functional relations, 
\ben
&&\hat  d(u) \hat g(u,v,m)+\hat  a(u) \hat n(u,v,m)=\tilde F(u,v)\phi(q^{-1}v^{-1})\hat  d(v)+\hat  \chi_{m-2}u^{-1} c(q u) \phi(q^{-1}v^{-1}),\\
&&\hat  a(u) \hat k(u,v,m)+\hat  d(u) \hat w(u,v,m)=-\tilde F(u,v)\phi(v)\hat   a(v)-\hat  \chi_{m-2}u^{-1} c(q u)\\
&&- \hat k(u,v,m)+\phi(q^{-1}u^{-1})   \hat w(u,v,m)=-G(u,v)b(q^{-2}v^{-2})+\hat  \r_{m-2}
\\
&&\phi(q^{-1}u^{-1})   \hat g(u,v,m)- \hat n(u,v,m)=\phi(q^{-1}v^{-1})\big(G(u,v)b(v^{2})-\hat  \r_{m-2}\big)\een
where
\ben
\hat  \chi_m=\chi_m\frac{\gamma(m-1)}{\gamma(m+1)},\quad \hat  \r_m=\rho_m\frac{\gamma(m-1)}{\gamma(m+1)},
\een
we are able to obtain,
\ben
&&\hat t_d(u,m)C(\bar u,m,M)=C(\bar u,m,M)\Big(h(u,\bar u)\hat a(u)\mathscr{\hat A}(u,m+2M)+f(u,\bar u)\hat d(u)\mathscr{\hat D}(u,m+2M)\Big)\\
&&\qquad\qquad +\sum_{i=1}^M\tilde{F}(u,u_i)C(\{u,\bar u_i\},m,M)\big(-\phi(u_i)h(u_i,\bar u_i)\hat a(u_i)\mathscr{\hat A}(u_i,m+2M)\nonumber\\
&&\qquad \qquad \qquad \qquad \qquad \qquad \qquad \qquad \qquad  +\phi(q^{-1}u^{-1}_i)f(u_i,\bar u_i)\hat d(u_i)\mathscr{\hat D}(u_i,m+2M)\big)\nonumber\\
&&\qquad\qquad +\hat \chi_mu^{-1} c(q u)\sum_{i=1}^MC(\{u,\bar u_i\},m,M)\big(-h(u_i,\bar u_i)\mathscr{\hat A}(u_i,m+2M)\nonumber\\
&&\qquad \qquad \qquad \qquad \qquad \qquad \qquad \qquad \qquad  +\phi(q^{-1}u_i^{-1})f(u_i,\bar u_i)\mathscr{D}(u_i,m+2M)\big)\nonumber
\een
and
\ben
&&\hat t_{ps}(u,m)C(\bar u,m,M)=C(\bar u,m,M)
\Big(-h(u,\bar u)\mathscr{\hat A}(u,m+2M)+\phi(q^{-1}u^{-1})f(u,\bar u)\mathscr{\hat D}(u,m+2M)\Big)\\
&&\qquad\qquad +\sum_{i=1}^M G(u,u_i)C(\{u,\bar u_i\},m,M)\big(-b(q^{-2}u_i^{-2})h(u_i,\bar u_i)\mathscr{\hat A}(u_i,m+2M)\nonumber\\
&&\qquad \qquad \qquad \qquad \qquad \qquad \qquad \qquad \qquad 
+b(u_i^2)\phi(q^{-1}u_i^{-1})f(u_i,\bar u_i)\mathscr{\hat D}(u_i,m+2M)\big)\nonumber\\
&&\qquad\qquad -\hat\r_m\sum_{i=1}^MC(\{u,\bar u_i\},m,M)\big(-h(u_i,\bar u_i)\mathscr{\hat A}(u_i,m+2M)\nonumber\\
&&\qquad \qquad \qquad \qquad \qquad \qquad \qquad \qquad \qquad +\phi(q^{-1}u_i^{-1})f(u_i,\bar u_i)\mathscr{\hat D}(u_i,m+2M)\big).\nonumber
\een
\end{rmk}


\section{Representation theory \label{S:Rep}}

We follow \cite{gauge} and construct the highest weight
vector by means of the covariant vector (\ref{coVec}).
For the choice of the gauge parameter,
\ben\label{alphahw}
\alpha=\alpha_{hw}= i q^{m_0+N} \frac{\tau  \mu}{\tilde\tau \tilde \mu},
\een
one can show that the vector
\bea\label{Dhwv}
|\Omega_{m_0}^N\rangle=\otimes_{i=1}^NX_i(v_i,m_0+i)
=X_1(v_1,m_0+1) \otimes \dots  \otimes X_N(v_N,m_0+N),
\eea
is the highest weight vector, such that the actions of the dynamical
operators at the point $m=m_0$ are given by,
\bea\label{actDhwv}
&&\mathscr{A}(u,m_0)|\Omega_{m_0}^N\rangle=u\, \tilde k^-(u) \Lambda(u)|\Omega_{m_0}^N\rangle,\\
\label{actDhwv1}&&\mathscr{D}(u,m_0)|\Omega_{m_0}^N\rangle=u \,\phi(q^{-1}u^{-1})\tilde k^-(q^{-1}u^{-1})
\Lambda(q^{-1}u^{-1})|\Omega_{m_0}^N\rangle,\\
\label{actDhwv2}&&\mathscr{C}(u,m_0)|\Omega_{m_0}^N\rangle=0.
\eea
where
\ben
 \Lambda(u) = \prod_{i=1}^{N}b(qu/v_i)b(quv_i).
\een

\begin{rmk}
The actions of the dynamical operators
are obtained in \cite{gauge} by introducing an
additional family of local gauged operators. We refer to the work
\cite{gauge} for more details on this point.
\end{rmk}

\begin{rmk}
For the other family of dynamical operators, $\{\hat{\mathscr{A}}(u,m),\hat{\mathscr{D}}(u,m)\}$, the gauge parameter
is fixed by,
\ben\label{betalw}
\beta=\beta_{lw}=i q^{-m_0-N} \frac{\tau\tilde \mu}{\tilde\tau  \mu}
\een
such that
\ben\label{Dlwv}
|\hat\Omega_{m_0}^N\rangle=\otimes_{i=1}^NY_i(v_i,m_0+2N-i)
=Y_1(v_1,m_0+2N-1)\otimes \dots \otimes Y_N(v_N,m_0+N),
\een
is the lowest weight vector satisfying,
\bea\label{ADlwv}
&&\mathscr{\hat A}(u,m_0+2N)|\hat\Omega_m^N\rangle=u \,\phi(q^{-1}u^{-1})\tilde k^-(qu)
\Lambda(q^{-1}u^{-1})|\hat\Omega_{m_0}^N\rangle\\
&&\mathscr{\hat D}(u,m_0+2N)|\hat\Omega_m^N\rangle=u\, \tilde k^-(u^{-1}) \Lambda(u)|\hat\Omega_{m_0}^N\rangle\\
&&\mathscr{B}(u,m_0+2N)|\hat\Omega_{m_0}^N\rangle=0.
\eea

\end{rmk}

\begin{rmk}\label{Rmkpart} 
If we fix at the same time both the parameters $\alpha=\alpha_{hw}$ and $\beta=\beta_{lw}$,
the highest and lowest weights vectors are related by,
\ben
\mathscr{B}\big(u_1,m_0+2(N-1)\big)\dots \mathscr{B}(u_N,m_0)|\Omega_{m_0}^N
\rangle=Z^N(\bar u| \bar v)|\hat\Omega_{m_0}^N\rangle
\een
and
\ben
\mathscr{C}\big(u_1,m_0+2\big)\dots \mathscr{C}(u_N,m_0+2N)|\hat \Omega_{m_0}^N\rangle=
\hat Z^N(\bar u, \bar v)|\Omega_{m_0}^N\rangle.
\een
where $Z^N(\bar u| \bar v)$ and $\hat Z^N(\bar u| \bar v)$ are the partition function
for the trigonometric solid-on-solid model with one reflecting end and
domain wall boundary conditions given explicitly in \cite{YCHHSZ11,gaugeKF}. In other words,
for such choice of the gauge parameters the dynamical creation and annihilation operators are
nilpotent, {\it i.e.},
\ben
B(\bar u,m_0+2(N+1),N+1)=
C(\bar u,m_0,N+1)=0.
\een
Moreover, this choice of gauge parameters maps the original problem
into a solid-on-solid model with left general and right diagonal boundaries.
\end{rmk}



\section{MABA for two general boundaries\label{S:MABA}}

We are now in position to implement the MABA for the Heisenberg XXZ spin
chain on the segment with two general boundaries.
In order to construct the Bethe vectors from the highest
weight vector and the dynamical $\mathscr{B}$ operator, we have to fix one of the gauge parameters
(\ref{alphahw}). We still have at our disposal one free gauge parameter,
which we use to bring the dynamical transfer matrix into a left lower and right upper dynamical form, and that allows us to follow the method introduced in \cite{Bel14} to implement the MABA.
This is the case for the choice of parameters,
\ben\label{gaugelowup}
\alpha=\alpha_{hw},\quad\quad\quad\quad \beta=\beta_{tl}(M)=-iq^{1-m_0-2M}\frac{\tilde\xi \tilde\kappa}{\xi\kappa},
\een
with $M$ some integer with values in $\{0,1,\dots,N\}$.
In this case, the coefficient of the operator $\mathscr{C}(u,m_0+2M)$ in the dynamical transfer matrix (\ref{t}) at the point $m_0+2M$
vanishes, \textit{i.e.},
\ben\label{tdlow}
&&t(u,m_0+2M)=t_d(u,m_0+2M) +u^{-1} c(qu)\Big( \zeta_{m_0+2M}\mathscr{B}(u,m_0+2M)-\delta_{m_0+2M}t_{ps}(u,m_0+2M)\Big).
\een

From the highest weight vector (\ref{Dhwv}) and the string of dynamical creation operators (\ref{SB}),
we can construct the vectors
\ben\label{Psi}
\Psi_{m_0}^{M}(\bar u)=B(\bar u,m_0+2M,M)|\Omega^N_{m_0}\rangle.
\een
Using the relation (\ref{Atd}) as well as the representation theory results (\ref{actDhwv},\ref{actDhwv1})
we obtain,
\ben\label{tdPsi}
&&t_d(u,m_0+2M)\Psi_{m_0}^{M}(\bar u)= \Lambda^M_{gd}(u,\bar u)\Psi_{{m_0}}^{M}(\bar u)\\
&&\qquad\qquad+\sum_{i=1}^M\big(\tilde F(u,u_i)E^M_{gd}(u_i,\bar u_i)+
\chi_{{m_0}+2M}\,u^{-1} c(q u)W^M(u_i,\bar u_i)\big)\Psi_{{m_0}}^{M}(\{u,\bar u_i\}),\nonumber
\een
with\footnote{Here and in the following the Bethe equations can be recovered from the corresponding eigenvalues
using the relation,
\ben
E^M(u_i,\bar u_i)=\lim_{u\to u_i}\Big(b(u_i/u)\Lambda^M(u,\bar u)\Big).
\nonumber
\een} 
\bea\label{Lamdg}
&& \Lambda_{gd}^M(u,\bar u)=\phi(u)\tilde k^+(u)\tilde k^-(u)\Lambda(u) f(u,\bar u)+\phi(q^{-1}u^{-1})\tilde k^+(q^{-1}u^{-1})\tilde k^-(q^{-1}u^{-1})\Lambda(q^{-1}u^{-1}) h(u,\bar u),
\\
&&E_{gd}^M(u_i,\bar u_i) =\phi(q^{-1}u^{-1}_i)\phi(u_i)\Big(\tilde k^+(u_i)\tilde k^-(u_i)\Lambda(u_i)f(u_i,\bar u_i)\label{Egd}\\
&&\qquad  \qquad  \qquad \qquad  \qquad
-\tilde k^+(q^{-1}u_i^{-1})\tilde k^-(q^{-1}u_i^{-1})\Lambda(q^{-1}u_i^{-1}) h(u_i,\bar u_i)\Big),\nonumber \een
and
\ben
W^M(u_i,\bar u_i)=
u_i \phi(q^{-1}u_i^{-1})\Big(\tilde k^-(u_i)\Lambda(u_i)f(u_i,\bar u_i)
-\tilde k^-(q^{-1}u_i^{-1})\Lambda(q^{-1}u_i^{-1})h(u_i,\bar u_i)\Big).
\een
Analogously, the action of $t_{ps}(u,{m_0}+2M)$ on (\ref{Psi}) is given by,
\ben\label{tpsPsi}
&& t_{ps}(u,{m_0}+2M)\Psi_{{m_0}}^{M}(\bar u)=\Lambda^M_{ps}(u,\bar u)\Psi_{{m_0}}^{M}(\bar u)\\
&&\qquad + \sum_{i=1}^M\big(G(u,u_i)b(q u_i^{2})E^M_{ps}(u_i,\bar u_i)
+\rho_{{m_0}+2M}W^M(u_i,\bar u_i)\big)\Psi_{{m_0}}^{M}(\{u,\bar u_i\})\nonumber
\een
with
\ben
 &&\Lambda^M_{ps}(u,\bar u)=u \phi(q^{-1}u^{-1})\Big(\tilde k^-(u)\Lambda(u)f(u,\bar u)-\tilde k^-(q^{-1}u^{-1})\Lambda(q^{-1}u^{-1})h(u,\bar u)\Big),\\
&&E^M_{ps}(u_i,\bar u_i)=u_i \phi(q^{-1}u_i^{-1})\Big( \phi(q^{-1}u_i^{-1})\tilde k^-(u_i)\Lambda(u_i)f(u_i,\bar u_i)+\phi(u_i)\tilde k^-(q^{-1}u_i^{-1})\Lambda(q^{-1}u_i^{-1})h(u_i,\bar u_i)\Big).
\een
Gathering the equations (\ref{tdPsi},\ref{tpsPsi}), it follows
that the full action of the dynamical transfer matrix (\ref{tdlow}) on (\ref{Psi}) is given by,
\ben\label{tlowPsi}
&& t(u,{m_0}+2M)\Psi_{{m_0}}^{M}(\bar u)=
\Lambda^M_{gd}(u,\bar u)\Psi_{{m_0}}^{M}(\bar u)+
\sum_{i=1}^M\tilde{F}(u,u_i)E^M_{gd}(u_i,\bar u_i)\Psi_{{m_0}}^{M}(\{u,\bar u_i\})\\
&&\qquad\qquad  +u^{-1} c(qu)\Big(\zeta_{{m_0}+2M}\mathscr{B}(u,{m_0}+2M)\Psi_{{m_0}}^{M}(\bar u)
-\delta_{{m_0}+2M}\Lambda^M_{ps}(u,\bar u)\Psi_{{m_0}}^{M}(\bar u) \nonumber\\
&&\qquad\qquad +\sum_{i=1}^M\big(\bar \chi_{{m_0}+2M}W^M(u_i,\bar u_i)-
\delta_{{m_0}+2M}G(u,u_i)b(q u_i^{2})E^M_{ps}(u_i,\bar u_i)\big)\Psi_{{m_0}}^{M}(\{u,\bar u_i\})\Big)\nonumber
\een
where we introduce,
\ben
&&\bar \chi_{m}=\chi_{m}-\delta_{m}\rho_{m}.
\een

Up to this point, we have basically an usual algebraic Bethe ansatz analysis.
The new feature,  which motivates the adjective {\it modified} for the ABA, is the presence of the modified creation operator $\mathscr{B}(u,{m_0}+2M)$
in the off-shell equation (\ref{tlowPsi}). For $M=N$ and $\bar u= \{u_1, \dots, u_N\}$, we can show from small
chain calculations\footnote{The unknown $\Lambda^N_{g}(u,\bar u)$ and $E^N_{g}(u_i,\bar u_i)$ are fixed from the direct resolution of the case $N=1$ and $N=2$ using symbolic calculation on Mathematica for generic boundary parameters and variables $\bar u$ and then verified for $N=3$ using arbitrary numerical parameters and variables.} that the vector $\mathscr{B}(u,{m_0}+2N)\Psi_{{m_0}}^{N}(\bar u)$
has an off-shell action given by,
\ben\label{offshellB}
&&u^{-1} c(qu)\zeta_{{m_0}+2N}\mathscr{B}(u,{m_0}+2N)\Psi_{{m_0}}^{N}(\bar u)
=
\Big(\Lambda^N_{g}(u,\bar u)+u^{-1} c(qu)\delta_{{m_0}+2N}\Lambda^N_{ps}(u,\bar u)\Big)\Psi_{m_0}^{N} (\bar u)\\
&&\nonumber \qquad\qquad\qquad +\sum_{i=1}^N
\Big(u^{-1} c(qu)\big(\delta_{{m_0}+2N}G(u,u_i)b(q u_i^{2})E^N_{ps}(u_i,\bar u_i)-\bar \chi_{{m_0}+2N}W^N(u_i,\bar u_i)\big)\\
\nonumber
&&\qquad\qquad\qquad\qquad\qquad\qquad\qquad+\tilde{F}(u,u_i)E^N_{g}(u_i,\bar u_i) \Big)\Psi_{m_0}^{N}(\{u,\bar u_i\})
\een
with
\bea
&& \Lambda_g^N(u,\bar u)=
-\kappa \tilde \kappa \tau  \tilde\tau
 \Big(\frac{ \kappa \tau}{\tilde \kappa  \tilde\tau}+\frac{\tilde \kappa  \tilde \tau}{ \kappa \tau}
 +\frac{\xi  \tilde \mu}{ \tilde \xi \mu} q^{N+1}+\frac{\tilde \xi \mu}{\xi \tilde \mu} q^{-N-1}\Big)
 c(u)c(q^{-1}u^{-1}) \Lambda(u)\Lambda(q^{-1}u^{-1})G(u,\bar u),\label{Lamg}\\
&&  E_g^N(u_i,\bar u_i) =\kappa \tilde \kappa \tau  \tilde\tau
 \Big(\frac{ \kappa \tau}{\tilde \kappa  \tilde\tau}+\frac{\tilde \kappa  \tilde \tau}{ \kappa \tau}
 +\frac{\xi  \tilde \mu}{ \tilde \xi \mu} q^{N+1}+\frac{\tilde \xi \mu}{\xi \tilde \mu} q^{-N-1}\Big)
\frac{ c(u_i)c(q^{-1}u_i^{-1}) }{b(q u_i^2)}\Lambda(u_i)\Lambda(q^{-1}u_i^{-1})G(u_i,\bar u_i)\label{Eg}.
\eea

The equation (\ref{offshellB}), together with (\ref{tlowPsi}), allows one to reach the main result
of this paper, the off-shell action of the
transfer matrix with two general boundaries given by,
\bea\label{tronBV}
&&t(u)\Psi_{m_0}^N(\bar u)=\Lambda^N(u,\bar u)\Psi_{m_0}^N(\bar u)+
\sum_{i=1}^N\tilde{F}(u,u_i)E^N(u_i,\bar u_i)\Psi_{m_0}^N(\{u,\bar u_i\})
\een
where
\ben
&&\Lambda^N(u,\bar u)=\Lambda^N_{gd}(u,\bar u)+\Lambda^N_g(u,\bar u),\quad E^N(u_i,\bar u_i) =E_{gd}^N(u_i,\bar u_i)+E_g^N(u_i,\bar u_i).\label{LamE}
\eea
with $\Lambda^N_{gd}(u,\bar u)$, $E_{g}^N(u_i,\bar u_i)$,  $\Lambda^N_{g}(u,\bar u)$ and
$E_{gd}^N(u_i,\bar u_i)$ given respectively by (\ref{Lamdg},\ref{Egd},\ref{Lamg},\ref{Eg})
and with the Bethe vector,
\ben\label{BVG}
\Psi_{{m_0}}^N(\bar u)=\mathscr{B}(u_{1},{m_0}+2(N-1) )\dots\mathscr{B}(u_N,{m_0})|\Omega^N_{{m_0}}\rangle
\een
for the gauge parameters (\ref{gaugelowup}) with $M=N$. 

 \begin{rmk} The eigenvalues (\ref{LamE}) correspond, up to a change of notation,
to the ones obtained in \cite{CYSW3, KMN14} and investigated at the thermodynamic
limit in \cite{termoXXZCao}. We remark that the new term in the eigenvalue expression,
characteristic of the modified T-Q Baxter equation introduced in \cite{CYSW1}, takes its algebraic
origin in the off-shell action of the creation operator (\ref{offshellB}).
\end{rmk}

\begin{rmk} The Bethe vector (\ref{BVG}) is constructed
from the dynamical $ \mathscr{B}$ operator with gauge parameters (\ref{gaugelowup}) which
correspond to bring the transfer matrix into a dynamical lower/upper triangular form.
It is also
possible to characterise this Bethe vector using
another set of gauge parameters which bring the  transfer matrix into a dynamical general/diagonal form, with 
\ben
\alpha=\alpha_{hw}, \quad \beta=\beta_{lw}.
\een
In this case both the highest and lowest weight vector do exist
at the same fixed $m=m_0$ and are related, see remark \ref{Rmkpart}. 
Using the relations between
the dynamical creation operator $\mathscr{B}(u,m)$ with dynamical parameters $\alpha$ and $\beta_{tl}$ and
the dynamical operators $\{\mathscr{A}(u,p),\mathscr{B}(u,p),\mathscr{C}(u,p),\mathscr{D}(u,p)\}$ with
dynamical parameters $\alpha$ and $\beta_{lw}$, given by
  \ben
&&  \mathscr{B}(u,m)|_{\beta_{tl}}=q^{m-p}\frac{(\alpha-q^{p+m}\beta_{tl})(\alpha-q^{2+p+m}\beta_{tl})}{(\alpha-q^{2+2m}\beta_{tl})(\alpha-q^{2p}\beta_{lw})}\mathscr{B}(u,p)|_{\beta_{lw}}\nonumber\\
  && +q^{3+m}\frac{(\alpha-q^{p+m}\beta_{tl})(q^m\beta_{tl}-q^{p}\beta_{lw})}{(\alpha-q^{2+2m}\beta_{tl})(\alpha-q^{2+2p}\beta_{lw})}t_{ps}(u,p)|_{\beta_{lw}}-q^{m+p}\frac{(q^p\beta_{lw}-q^{m}\beta_{tl})(q^p\beta_{lw}-q^{2+m}\beta_{tl})}{(\alpha-q^{2+2m}\beta_{tl})(\alpha-q^{2p}\beta_{lw})}\mathscr{C}(u,p)|_{\beta_{lw}}\nonumber
  \een
  which can be obtained using (\ref{DyntoK}),
we can project  (\ref{BVG})
in this new basis of dynamical operators with $\alpha=\alpha_{hw}$ and $\beta=\beta_{lw}$, namely
 \ben
 \Psi_{m_0}^N(\bar u)=\sum_{i=0}^N\sum_{\bar u \Rightarrow \{\bar u_{\so},\bar u_{\st}\}}W_{N-i}^N(\bar u_{\so}|\bar u_{\st})\bar\Psi^i_{m_0}(\bar u_{\st})
 \een
with 
 \ben
 &&\bar\Psi^i_{m_0}(\{u_{i+1},\dots,u_N\})=\mathscr{B}(u_{i+1},m_0+2(N-i) )\dots\mathscr{B}(u_N,m_0)|\Omega^N_{m_0}\rangle.
 \een
and where the second sum
corresponds to each splitting of the set $\bar u$ into subsets $\bar u_{\so}$ and
$\bar u_{\st}$ with $\#\bar u_{\st}=N-i$ and where the elements in every subset are
ordered in such a way that the sequence of their subscripts is strictly increasing. 
The explicit coefficients $W_{N-i}^N(\bar u_{\so}|\bar u_{\st})$ can be fixed by
the expansion and will be considered elsewhere.
\end{rmk}


\begin{rmk} The MABA can be similarly performed using the $\mathscr{C}(u,m)$
operators and the lowest weight vector  (\ref{Dlwv})  to construct the Bethe Vector.
In this case, the lowest weight vector fixes the gauge parameter,
\ben\label{PC1}
\beta=\beta_{lw}=i q^{-m_{0}-N} \frac{\tau\tilde \mu}{\tilde\tau  \mu},
\een
and, by means of the choice,
\ben\label{PC2}
\alpha=\alpha_{tu}(\hat M)=-iq^{1+m_0+2(N-\hat M)}\frac{\xi \tilde{\kappa}}{\tilde\xi\kappa}
\een
the transfer matrix acquires a dynamical upper/lower structure at the point $m_0+2(N-\hat M)$,
namely,
\ben\label{tdup}
&&t(u,m_0+2(N-\hat M))=\hat t_d(u,m_0+2(N-\hat M)) \nonumber\\
&&-u^{-1} c(qu)\Big(\tilde{\zeta}_{m_0+2(N-\hat M)}\mathscr{C}(u,m_0+2(N-\hat M))+
\delta_{m_0+2(N-\hat M)-2}\hat{t}_{ps}(u,m_0+2(N-\hat M))\Big).
\een
We introduce the vector,
\ben
&& \hat \Psi_{{m_0}}^{\hat  M}(\bar u)=C(\bar u,{m_0}+2(N-\hat M),\hat M)|\hat \Omega^N_{{m_0}}\rangle
\een
such that at the point $m_0+2(N-\hat M)$, we have the off-shell action,
\ben\label{Off-C}
&& t(u,{m_0}-2 \hat M+2N)\hat\Psi_{{m_0}}^{\hat M}(\bar u)=
\left(
\hat\Lambda^{\hat M}_d(u,\bar u)+
u^{-1} c(qu)\delta_{{m_0}-2M+2N-2}\hat\Lambda^{\hat M}_{ps}(u,\bar u)\right)\hat\Psi_{{m_0}}^{M}(\bar u)
\\
&&\qquad\qquad +
\sum_{i=1}^{\hat M}
\left[
\tilde{F}(u,u_i)\hat E^{\hat M}_d(u_i,\bar u)+u^{-1}c(qu)
\left(
\hat{\bar{\chi}}_{{m_0}-2{\hat M}+2N}\hat W^{\hat M}(u_i,\bar u_i)\right.\right. \nonumber\\
&& \qquad\qquad+ \left. \left.
\delta_{{m_0}-2{\hat M}+2N-2}G(u,u_i)b(q^{-1} u_i^{-2})\hat E^{\hat M}_{ps}(u_i,\bar u_i)
\right)
\right]\hat \Psi_{{m_0}}^{\hat M}(\{u,\bar u_i\})\nonumber\\
&&\qquad\qquad 
-u^{-1} c(qu)\tilde\zeta_{{m_0}-2{\hat M}+2N}\mathscr{C}(u,{m_0}-2{\hat M}+2N)\hat\Psi_{{m_0}}^{\hat M}(\bar u)\nonumber
\een
where we define
\ben
\hat{\bar{\chi}}_m=\hat \chi_m-\delta_{m-2}\hat \rho_m
\een
and with the functions,
\bea\label{BonBVC}
&& \hat\Lambda_{gd}^{\hat M}(u,\bar u)=\phi(u)\tilde k^+(u^{-1})\tilde k^-(u^{-1})\Lambda(u) f(u,\bar u)+\phi(q^{-1}u^{-1})\tilde k^+(qu)\tilde k^-(qu)\Lambda(q^{-1}u^{-1}) h(u,\bar u),
\\
&&\hat E_{gd}^{\hat M}(u_i,\bar u_i) =\phi(q^{-1}u^{-1}_i)\phi(u_i)\Big(\tilde k^+(u_i^{-1})\tilde k^-(u_i^{-1})\Lambda(u_i)f(u_i,\bar u_i)\label{hEgd}\\
&&\qquad  \qquad  \qquad \qquad  \qquad  \qquad  \qquad \qquad  \qquad    -\tilde k^+(qu_i)\tilde k^-(qu_i)\Lambda(q^{-1}u_i^{-1}) h(u_i,\bar u_i)\Big),\nonumber \een
\ben
 && \hat \Lambda^{\hat M}_{ps}(u,\bar u)=u \phi(q^{-1}u^{-1})\Big(\tilde k^-(u^{-1})\Lambda(u)f(u,\bar u)-\tilde k^-(qu)\Lambda(q^{-1}u^{-1})h(u,\bar u)\Big),\\
&& \hat E^{\hat M}_{ps}(u_i,\bar u_i)=u_i \phi(q^{-1}u_i^{-1})\Big( \phi(q^{-1}u_i^{-1})\tilde k^-(u_i^{-1})\Lambda(u_i)f(u_i,\bar u_i)+\phi(u_i)\tilde k^-(qu_i)\Lambda(q^{-1}u_i^{-1})h(u_i,\bar u_i)\Big),\\
&& \hat W^{\hat M}(u_i,\bar u_i)=
u \phi(q^{-1}u_i^{-1})\Big(\tilde k^-(u_i^{-1})\Lambda(u_i)f(u_i,\bar u_i)-\tilde k^-(qu_i)\Lambda(q^{-1}u_i^{-1})h(u_i,\bar u_i)\Big).\een
For $\hat M=N$, we can conjecture that 
\ben
&&-u^{-1} c(qu)\tilde\zeta_{{m_0}}\mathscr{C}(u,{m_0})\hat\Psi_{{m_0}}^{N}(\bar u)
=
\Big(\hat\Lambda^N_{g}(u,\bar u)-u^{-1} c(qu)\delta_{{m_0}-2}
\hat\Lambda^N_{ps}(u,\bar u)\Big)\hat\Psi_{m_0}^{N} (\bar u)\\
&&\nonumber \qquad\qquad\qquad +
\sum_{i=1}^N\Big(-u^{-1} c(qu)\big(\delta_{{m_0}-2}G(u,u_i)b(q^{-1} u_i^{-2})\hat E^N_{ps}(u_i,\bar u_i)+
\hat{\bar{\chi}}_{{m_0}}\hat W^N(u_i,\bar u_i)\big)\\
\nonumber
&&\qquad\qquad\qquad\qquad\qquad\qquad\qquad
+\tilde{F}(u,u_i)\hat E^N_{g}(u_i,\bar u_i) \Big)\hat\Psi_{m_0}^{N}(\{u,\bar u_i\}),
\een
with
\bea
&& \hat \Lambda_g^N(u,\bar u)=
-\kappa \tilde \kappa \tau  \tilde\tau
 \Big(\frac{ \kappa \tau}{\tilde \kappa  \tilde\tau}+\frac{\tilde \kappa  \tilde \tau}{ \kappa \tau}
 +\frac{ \tilde\xi  \tilde \mu}{ \xi \mu} q^{-N-1}+\frac{ \xi \mu}{ \tilde\xi \tilde \mu} q^{N+1}\Big)
 c(u)c(q^{-1}u^{-1}) \Lambda(u)\Lambda(q^{-1}u^{-1})G(u,\bar u),\label{Lamgc}\\
&& \hat E_g^N(u_i,\bar u_i) =\kappa \tilde \kappa \tau  \tilde\tau
 \Big(\frac{ \kappa \tau}{\tilde \kappa  \tilde\tau}+\frac{\tilde \kappa  \tilde \tau}{ \kappa \tau}
 +\frac{ \tilde\xi  \tilde \mu}{  \xi \mu} q^{-N-1}+\frac{ \xi \mu}{ \tilde\xi \tilde \mu} q^{N+1}\Big)
\frac{ c(u_i)c(q^{-1}u_i^{-1}) }{b(q u_i^2)}\Lambda(u_i)\Lambda(q^{-1}u_i^{-1})G(u_i,\bar u_i),\label{Egc}
\eea
which gives the final result for the dynamical $\mathscr{C}$ operator, namely,
\bea\label{tronBVC}
&&t(u)\hat \Psi_{m_0}^N(\bar u)=\hat \Lambda^N(u,\bar u)\hat \Psi_{m_0}^N(\bar u)+
\sum_{i=1}^N\tilde{F}(u,u_i)\hat{E}^N(u_i,\bar u_i)\hat \Psi_{m_0}^N(\{u,\bar u_i\})\\
&&\hat \Lambda^N(u,\bar u)=\hat \Lambda^N_{gd}(u,\bar u)+\hat \Lambda^N_g(u,\bar u),\quad \hat E^N(u_i,\bar u_i) =
\hat E_{gd}^N(u_i,\bar u_i)+\hat E_g^N(u_i,\bar u_i),\label{LamEC}
\eea
where the Bethe vector is given by
\ben
\hat \Psi_{{m_0}}^N(\bar u)=\mathscr{C}(u_{1},{m_0}+2)\dots\mathscr{C}(u_N,{m_0}+2N)|\hat \Omega^N_{{m_0}}\rangle
\een
with gauge parameters (\ref{PC1}) and (\ref{PC2}) with $\hat M=N$.
Due to the symmetry of the parametrisation (\ref{NpKm},\ref{NpKp}),
the resulting on-shell eigenvalues are the same of those ones (\ref{LamE}) which are obtained from
the dynamical $\mathscr{B}$ operator in the off-shell case. 

\end{rmk}

\section{Some limiting cases \label{S:Lim}} 

We consider, from the general
result of the previous section,
two special limits of the XXZ spin chain on the segment.
Firstly, we consider the limit where the right boundary is
right upper triangular and, next, cases where there exist constraints between
left and right boundary couplings.

\subsection{Limit to the right triangular boundary} 
Here we consider the limit $\tilde \tau \to0$ from the
previous results.
In this limit, the $K^--$ matrix (\ref{Km}) becomes upper triangular
and the off-shell action for the transfer matrix (\ref{tronBV}) is given by,
\ben\label{tPHI}
&&t(u)\Phi_{{m_0}}^N(\bar u)=\Lambda_{up}^N(u,\bar u)\Phi_{{m_0}}^N(\bar u)+
\sum_{i=1}^N\tilde{F}(u,u_i)\text{E}_{up}^N(u_i,\bar u_i)\Phi_{{m_0}}^N(\{u,\bar u_i\})
\een
with
\ben
\Lambda_{up}^N(u,\bar u)&=&\Lambda^N_d(u,\bar u)+\Lambda^N_{gup}(u,\bar u),\quad 
\text{E}_{up}^N(u_i,\bar u_i)=\text{E}_d^N(u_i,\bar u_i)+\text{E}_{gup}^N(u_i,\bar u_i)
\een
where
\ben\label{limtLd}
&&\Lambda^N_d(u,\bar u)=\phi(u)\tilde k^+(u)\big(u k^-(u)\big)\Lambda(u)f(u,\bar u)\\
&&\qquad \qquad\qquad \qquad +\phi(q^{-1}u^{-1})\tilde k^+(q^{-1}u^{-1})\big(q^{-1}u^{-1}k^-(q^{-1}u^{-1})\big)\Lambda(q^{-1}u^{-1})h(u,\bar u),\nonumber\\
&& \label{limtEd} \text{E}^N_d(u_i,\bar u_i)=\phi(q^{-1}u_i^{-1})\phi(u_i)\Big(\phi(u_i)\tilde k^+(u_i)\tilde k^+(u)\big(u_i k^-(u_i)\big)\Lambda(u_i)f(u_i,\bar u_i)\\
&&\qquad \qquad\qquad \qquad \qquad -\tilde k^+(q^{-1}u_i^{-1})\big(q^{-1}u_i^{-1}k^-(q^{-1}u_i^{-1})\big)\Lambda(q^{-1}u_i^{-1})h(u_i,\bar u_i)\Big).\nonumber
\een
and
\ben
&&\Lambda_{gup}^N(u,\bar u)=i\kappa\tilde\kappa\Big(\nu_-q^{-N-1}\frac{\tilde  \xi}{ \xi}+i\frac{\kappa }{\tilde \kappa}\tau^2\Big)c(u)c(q^{-1}u^{-1})\Lambda(u)\Lambda(q^{-1}u^{-1})G(u,\bar u),\\
&&\text{E}^N_{gup}(u_i,\bar u_i)=-i\kappa\tilde\kappa\Big(\nu_-q^{-N-1}\frac{\tilde  \xi}{ \xi}+i\frac{\kappa}{\tilde \kappa} \tau^2\Big)\frac{c(u_i)c(q^{-1}u_i^{-1})}{b(q u_i^2)}\Lambda(u_i)\Lambda(q^{-1}u_i^{-1})G(u_i,\bar u_i).
\een
The Bethe vector is given by
\ben
\Phi_{m_0}^N(\bar u)=\mathscr{B}(u_{1},m_0+2(N-1) )\dots\mathscr{B}(u_N,m_0)|\Omega\rangle
\een
and with 
\ben
\beta=\beta_{tl}=-iq^{1-m_0-2N}
\frac{\tilde\kappa\tilde\xi}{\kappa \xi}.
\een
Let us remark that the parameter $\alpha_{hw}$ does not appear in the off-shell action (\ref{tPHI}).
Indeed, this fact is expected since the general off-shell formula (\ref{tronBV})
depends explicitly on $\alpha_{hw}$ only through the highest
weight vector (\ref{Dhwv}). The dynamical $\mathscr{B}$ operator is independent of $\alpha_{hw}$ and
the transfer matrix, invariant by any gauge transformation,
is considered to be in the form (\ref{tr}) for the limit.
Thus, to obtain the limit, we only need to consider the leading term
of the dynamical highest weight vector (\ref{Dhwv}), namely
\ben
\text{ lim}_{\tilde \tau \to 0}|\Omega_{m_0}^N\rangle \sim \tilde \tau^{-N} |\Omega\rangle
\een
which gives the highest weight vector of the diagonal case \cite{Skl88}, given in appendix (\ref{rep}).
Moreover, we use
the limit of the $\tilde k^-(u)$ function,
\ben\label{limkm}
\text{ lim}_{\tilde \tau \to 0}\Big(\tilde k^-(u)\Big)= u k^-(u)
\een
as well as the limit of the constant coefficient in the new term of the eigenvalue (\ref{Lamgc}),
\ben\label{limtcoef}
\text{ lim}_{\tilde \tau \to 0}\Big(-\kappa \tilde \kappa \tau  \tilde\tau
 \Big(\frac{ \kappa \tau}{\tilde \kappa  \tilde\tau}+\frac{\tilde \kappa  \tilde \tau}{ \kappa \tau}
 +\frac{\xi  \tilde \mu}{ \tilde \xi \mu} q^{N+1}+\frac{\tilde \xi \mu}{\xi \tilde \mu} q^{-N-1}\Big)\Big)=
 i\kappa\tilde\kappa\Big(\nu_-q^{-N-1}\frac{\tilde  \xi}{ \xi}+i\frac{\kappa }{\tilde \kappa}\tau^2\Big).
\een 

\begin{rmk} 
This result can also be obtained directly from the MABA.
It is important to consider this fact since, rather than a simple limit, it allows one
to emphasise some characteristics which are hidden in the generic case. Firstly,
we put the transfer matrix into a modified diagonal form, {\it i.e.} with only $\mathscr{A}$ and $\mathscr{D}$ operators \cite{Bel14}, at the point $m_0+2M$ 
\bea\label{tgup}
t(u)=t_d(u,m_0+2M)=\tilde a(u)\mathscr{A}(u,m_0+2M)+\tilde d(u)\mathscr{D}(u,m_0+2M),
\eea
where we fix the gauge parameters to be 
\ben\label{albetd}
 \alpha= \alpha_d(M)=-i \frac{\tilde\kappa\xi}{\kappa \tilde \xi}q^{m_0+2M+1},
 \quad \beta= \beta_d(M)=-i \frac{\tilde\kappa \tilde\xi}{\kappa  \xi}q^{-m_0-2M+1},
 \een 
where $M$ is an integer in $\{0,1,\dots,M\}$.
From the explicit relation between the initial operators (\ref{KO}) and the dynamical one
(\ref{odyn}) given in (\ref{DyntoK}) and the action of the initial
operators on  the highest weight vector (\ref{Avac}) we can find,
\bea \label{actdynvacA}
&&\mathscr{A}(u,{m_0})|\Omega\rangle= u^2 k^+(u)\Lambda(u)|\Omega\rangle+\mathscr{B}(u,{m_0}-2)|\Omega\rangle\\ \label{actdynvacD}
&&\mathscr{D}(u,{m_0})|\Omega\rangle=\phi(q^{-1}u^{-1})k^+(q^{-1}u^{-1})\Lambda(q^{-1}u^{-1})|\Omega\rangle- \phi(u)\mathscr{B}(u,{m_0}-2)|\Omega\rangle.
\eea
Such actions are called modified, due to the presence of an off-diagonal term,
and were already pointed out in \cite{BC13,Bel14}. They are a new feature which can appear in the context of the MABA.
Then, using the string of $\mathscr{B}$ operators and the highest weight vector (\ref{Om}), we introduce the vector
\ben\label{BVdynM}
\Phi_{m_0}^M(\bar u)=B(\bar u,{m_0}+2M,M)|\Omega\rangle,
\een
and compute, from the commutation relations (\ref{AonSB},\ref{DonSB})
as well as from the modified off-diagonal action (\ref{actdynvacA},\ref{actdynvacD}),
the off-shell action of the modified transfer matrix (\ref{tgup}), given by
\ben
&&t(u)\Phi_{{m_0}}^M(\bar u)=
\Lambda^M_d(u,\bar u)\Phi_{{m_0}}^M(\bar u)+
\sum_{i=1}^M\tilde{F}(u,u_i)\text{E}^M_d(u_i,\bar u_i)\Phi_{{m_0}}^M(\{u,\bar u_i\})\\
&&\qquad \qquad\qquad \qquad+\kappa^2\gamma_{{m_0}-1}q^{-1}u^{-1}c(qu)\mathscr{B}(u,{m_0}+2M-2)\Phi_{{m_0}-2}^M(\bar u)\nonumber
\een
with $\Lambda^M_d(u,\bar u)$ and $ \text{E}^M_d(u_i,\bar u_i)$ are given by (\ref{limtLd},\ref{limtEd}) with $N\to M$
and where we use the following identities to simplify the term $\mathscr{B}(u,m-2)\Phi_{m-2}^M(\bar u)$,
\ben
&&\sum_{i=1}^{M}\Big(g(u,u_i,m)f(u_i,\bar u_i)-w(u,u_i,m)h(u_i,\bar u_i)\phi(u_i)\Big)=\frac{\gamma_{m-2M+1}}{\gamma_{m+1}}-f(u,\bar u),\\
&&\sum_{i=1}^{M}\Big(n(u,u_i,m)f(u_i,\bar u_i)-k(u,u_i,m)h(u_i,\bar u_i)\phi(u_i)\Big)=-\phi(u)\big(\frac{\gamma_{m-2M+1}}{\gamma_{m+1}}-h(u,\bar u)\big).
\een
For $M=N$, the action of the dynamical creation operator has an off-shell structure that we conjecture to be of the form
\ben
&&\quad  \kappa^2\gamma_{{m_0}-1}q^{-1}u^{-1}c(qu)\mathscr{B}(u,{m_0}-2+2N)\Phi_{{m_0}-2}^N(\bar u)=\\
&&\qquad \qquad \qquad \Lambda^N_{gup}(u,\bar u)
\Phi_{{m_0}}^N(\bar u)+\sum_{i=1}^N\tilde{F}(u,u_i)\text{E}^N_{gup}(u_i,\bar u_i)\Phi_{{m_0}}^N(\{u,\bar u_i\}).\nonumber
\een
In particular we note that $\beta_d(N)=\beta_{tl}$ which corresponds to the result we obtained by limit from the general boundary case.
\end{rmk}

\begin{rmk} We can also obtain by limit the cases with two upper triangular boundaries and  with lower and upper 
triangular boundaries considered in \cite{Bel14}\footnote{At this point, and only in this remark, we remove the square on the
off-diagonal boundary parameters to fit with \cite{Bel14}.}.
For the former, we consider the limit $\kappa \to 0$  and $\tilde \tau \to 0$ of the $\tilde k^-(u)$ function (\ref{limkm}), of the $\tilde k^+(u)$ function 
\ben
\text{ lim}_{ \kappa \to 0}\Big(\tilde k^+(u)\Big)= u^{-1} k^+(u),
\een
and of the parameters
\ben
&&\text{ lim}_{ \kappa \to 0, \tilde\tau \to 0}\, \Big(\delta_{m_0+2M}\Big)=\text{ lim}_{ \kappa \to 0,\tilde \tau \to 0}\Big(\zeta_{m_0+2M}\Big)=0,\\
&&\text{ lim}_{ \kappa \to 0, \tilde\tau \to 0}\Big(\tilde\zeta_{m_0+2M}\Big)=\text{ lim}_{\kappa \to 0, \tilde\tau \to 0}\Big(\chi_{m_0+2M}\Big)=0,\\
&&\text{ lim}_{\tilde \kappa \to 0}\Big(\beta_{tl}(M)\Big)\to q^{1-m_0-2M} \frac{\tilde\kappa}{\epsilon_-},
\een
on the off-shell action (\ref{tlowPsi}). For $m_0=-2M$, this allows to recover the off-shell action given by equation (4.12)  in \cite{Bel14}. The form of the vectors $Y$ and $\tilde Y$ behave, up to a factor that changes $\tilde F$ to $F$, as  
\ben 
Y_{up}(u,m)=\left(\begin{array}{c}
   \frac{\kappa} {\epsilon_-} q^{m+1} u^{-1},  \\
       1
      \end{array}
\right),\quad \tilde Y_{up}(u,m)=\left(
  - \frac{\kappa} {\epsilon_-} q^{m+1} u^{-1} ,\quad 
       1
\right).
\een
To recover the notation of \cite{Bel14} we have to choose the vectors $X$ and $\tilde X$ to be
\ben
X_{up}(u,m)=\left(\begin{array}{c}
  1 \\
        0
      \end{array}
\right),\quad
\tilde X_{up}(u,m)=\left(0, \quad 1
\right).
\een
This is allowed from the invariance of the transfer matrix  by any gauge transformation.  
For the latter, we took the limit $\tilde \kappa \to 0$ of the coefficient (\ref{limtcoef}), of the function $\tilde k^+(u)$
\ben
\text{ lim}_{\tilde \kappa \to 0}\Big(\tilde k^+(u)\Big)= u^{-1} k^+(u),
\een 
and of the gauge parameter
\ben
\text{ lim}_{\tilde \kappa \to 0}\Big(\beta_{tl}(M)\Big)= 0
\een
in the off-shell action (\ref{tPHI}) in order to recover the equation (5.9) of \cite{Bel14}.
The form of the vectors $Y$ and $\tilde Y$ behave, up to a factor that change $\tilde F$ to $F$, as  
\ben
Y_{lo}(u,m)=\left(\begin{array}{c}
  0 \\
       1
      \end{array}
\right),\quad
\tilde Y_{lo}(u,m)=\left(
  0,\quad 
       1
\right).
\een
To recover the notation of \cite{Bel14} we have to choose the vectors $X$ and $\tilde X$ to be
\ben
X_{lo}(u,m)=\left(\begin{array}{c}
  1 \\
         \frac{ \kappa} {\epsilon_-} q^{m-1} u
      \end{array}
\right),\quad
\tilde X_{lo}(u,m)=\left(
-     \frac{ \kappa} {\epsilon_-} q^{m-1} u, \quad 1
\right).
\een
While the expressions (\ref{Scal-prod}) for scalar products and
(\ref{close}) for the closure relation are the same,
the intertwining relations with the $R-$matrix (\ref{RXY},\ref{RtXtY}) are simpler.
In fact, in these cases all the functions $\gamma$ are set to the unity.
For this reason, the commutation relations among the dynamical
operators will depend on
the dynamical integer $m$ only through the operators, see \cite{Bel14}. 
\end{rmk}

\subsection{Limit to the cases with constraints between right and left boundary} 
Let us recover previous results in the literature. From the off-shell equations
(\ref{tlowPsi}) and (\ref{Off-C}), we note that special cases can be obtained
by setting,
\ben\label{conditionB}
 \zeta_{{m_0}+2M}=\delta_{{m_0}+2M}=\chi_{{m_0}+2M}=0
\een
and
\ben\label{conditionC}
 \tilde{\zeta}_{{m_0}-2\hat M+2N}=\delta_{{m_0}-2\hat M+2N-2}=\hat{\chi}_{{m_0}-2\hat M+2N}=0.
\een
In fact, if relations (\ref{conditionB}) and (\ref{conditionC}) are satisfied,
all off-diagonal contributions to the off-shell actions (\ref{tlowPsi}) and (\ref{Off-C}) disappear,
remaining only the usual off-shell diagonal terms, allowing the usual Bethe ansatz execution.
The constraints between the boundary parameters
which arise from (\ref{conditionB}) and (\ref{conditionC}) can be
written as, respectively,
\ben\label{constraintB}
 -\frac{\tilde{\kappa}\tilde{\mu}\tilde{\tau}\xi}
 {\kappa\mu\tau\tilde{\xi}}q^{1+2M-N}=1
\een
and
\ben\label{constraintC}
 -\frac{\kappa\xi\tau\tilde{\mu}}
 {\tilde{\kappa}\tilde{\tau}\tilde{\xi}\mu}q^{N-1-2\hat M}=1.
\een
If we require the relation between $M$ and $\hat M$ to be,
\ben
 M+\hat M=N-1
\een
and use the parameterisation,
\ben
 &&\tau=
 e^{\frac{\theta_{-}}{2}}\sqrt{\frac{\kappa_{-}}{2}}
 ,\quad 
 \tilde{\tau}=
 e^{-\frac{\theta_{-}}{2}}\sqrt{\frac{
 \kappa_{-}
 }{2}}, \quad
 \kappa=ie^{-\frac{\theta_{+}}{2}}\sqrt{\frac{\kappa_{+}}{2}}
 , \quad
 \tilde{\kappa}=ie^{\frac{\theta_{+}}{2}}\sqrt{\frac{\kappa_{+}}{2}},\nonumber\\
 &&
 \mu=-ie^{-\alpha_{-}},\quad
 \tilde{\mu}=-e^{\beta_{-}},\quad
 \xi=ie^{\alpha_{+}} ,\quad
 \tilde{\xi}=- e^{-\beta_{+}},
\een
we find that the constraints (\ref{constraintB}) and (\ref{constraintC}) reduce
to the ones obtained in \cite{TQ2}, namely
\ben
\alpha_{-}+\beta_{-}+\alpha_{+}+\beta_{+}=\pm(\theta_{-}-\theta_{+})+\eta k,\quad\quad k=N-2M-1.
\een
We observe that, in this case, the gauge transformation used to construct
the Bethe vector from the dynamical $\mathscr{B}$ operator is not the same
that for the one used to build
the Bethe vector from the dynamical $\mathscr{C}$ operator. On the order hand, if $M$ and $\hat M$ are related
by
\ben
M+\hat M=N,
\een
with the parameterisation
\ben
 &&\tau=
 e^{-\frac{\tau_{0}}{2}}\sqrt{\frac{1}{2}}
 ,\quad 
 \tilde{\tau}=
 ie^{\frac{\tau_{0}}{2}}\sqrt{\frac{
 1
 }{2}}, \quad
 \kappa=e^{\frac{\bar\tau}{2}}\sqrt{\frac{1}{2}}
 , \quad
 \tilde{\kappa}=ie^{-\frac{\bar\tau}{2}}\sqrt{\frac{1}{2}},\nonumber\\
 &&
 \mu=-ie^{-\delta},\quad
 \tilde{\mu}=-e^{-\zeta},\quad
 \xi=-ie^{-\bar\delta} ,\quad
 \tilde{\xi}=- e^{-\bar\zeta},
\een
we recover the constraints in \cite{gaugeKF} and we 
further observe, in this case, that the gauge parameters $\alpha$ and $\beta$
are the same for both the $\mathscr{B}$ and $\mathscr{C}$ cases.
These values of $\alpha$ and $\beta$ correspond to the ones required
to have highest and lowest weight vector (see remark \ref{Rmkpart}).


\section{Conclusion} \label{S:Conc}


In this work, we have constructed the Bethe vector of the Heisenberg XXZ spin chain on the 
segment with generic boundary parameters by means of the MABA approach. The off-shell action of the transfer matrix on
the Bethe vector is also given. The construction follows from the conjecture of
an specific off-shell structure for the action of 
the so-called modified creation operator on the Bethe vector.
\vspace{0.2cm}

From the general result, we obtain by limit the case
of the Heisenberg XXZ spin chain on the segment with left generic and right upper triangular
boundary parameters, considered in \cite{niccoli2} from another approach. Also, we recover the limiting case
where the left and right boundary parameters are related by constraints \cite{TQ2,gaugeKF}. 

\vspace{0.2cm}

The proof of the conjecture of the off-shell action of the modified creation operator on the Bethe vector,
which is the key feature of the MABA, remains to be done. It could be handled in different ways,
following for instance \cite{Cra14}, whose proof is based on the analytical
properties of the modified creation operator and in the fact that such operator is invertible.
Other possibilities would be to
consider the {\it off-diagonal Bethe ansatz} method presented in \cite{CYSW4} or by means of
an explicit link with the SOV approach \cite{FKN}.

\vspace{0.2cm}

Our results can be used to consider the scalar product of the Bethe vector. This is
a crucial step to consider the correlation functions of the Heisenberg XXZ spin chain on the 
segment with generic boundary parameters, within the algebraic Bethe ansatz framework.
This will extend the known results for the Heisenberg XXZ spin chain on the circle \cite{KMT},
on the segment with diagonal boundary \cite{KKMNST} or with general constrained boundary \cite{YCHHSZ11} conditions.
A key step will be to find a simple realisation of the  scalar product between an
off-shell and an on-shell Bethe vector, if possible, in terms of an unique determinant.

\vspace{0.2cm}

Finally, it is interesting to extend the MABA to other models without $U(1)$ symmetry,
in particular for finite models with higher spin and higher rank symmetry algebra.
In addition, it should be interesting to consider models
with an infinite dimensional Hilbert space, such as spin
chains with defect or the Bose gas on the segment or on the half-line.

\vspace{0.2cm}

{\bf Note added:} After this work was completed, we became aware of the paper \cite{CYSW5},
where the Bethe vector, in a different parametrisation, is obtained by another method.

\vspace{0.2cm}

{\bf Acknowledgement:} 
We thank J. Avan, P. Baseilhac, N. Cramp\'e, A. Lima-Santos and V. Pasquier for discussions.
S.B. is  partially supported by Sao Paulo Research Foundation (FAPESP), grant \# 2014/09832-1
and thanks the Departamento de F\'{\i}sica of the Universidade Federal de S\~ao Carlos for hospitality,
where part of this work has been done. 
R.A.P. is supported by FAPESP, grant \# 2014/00453-8.

\appendix



\section{Functions and commutation relations\label{App:Func}}
We use the following functions along the text,
\ben\label{fonctions1}
&&b(u)=\frac{u-u^{-1}}{q-q^{-1}},
\quad k^-(u) =\nu_-u+\nu_+u^{-1},
\quad k^+(u) =\epsilon_+u+\epsilon_-u^{-1},\quad  c(u)=u^2-u^{-2},\\
&&\tilde k^-(u)=
i \tilde\tau\tau ( \mu u +\mu^{-1} u^{-1})(\tilde\mu^{-1}u + \tilde\mu u^{-1}), \quad
\tilde k^+(u)=
i \tilde\kappa\kappa   ( \tilde \xi u +\tilde \xi^{-1} u^{-1})(\xi^{-1}u + \xi u^{-1}), \\
&&\phi(u)= \frac{b(q^2u^2)}{b(qu^2)},
\quad G(u,v)=\frac{1}{b(u/v)b(quv)}, \quad F(u, v)=G(u,v)\frac{b(q^2 u^2)}{\phi(v)}, \quad \tilde F(u,v)=(v/u) F(u,v), \\
&&f(u,v)= \frac{b(qv/u)b(uv)}{b(v/u)b(quv)}\ ,
\quad g(u,v)= \frac{\phi(q^{-1}v^{-1})}{b(u/v)}, \quad w(u,v)= -\frac{1}{b(quv)},\\
 &&h(u,v)= \frac{b(q^2uv)b(qu/v)}{b(quv)b(u/v)},\quad k(u,v)= \frac{\phi(u)}{b(v/u)}, \quad
n(u,v)= \frac{\phi(u)\phi(q^{-1}v^{-1})}{b(quv)},\\
&&s(u,v)=\frac{\phi(q^{-1}u^{-1})}{b(v/u)b(q v^2)},
\quad x(u,v)=\frac{\phi(q^{-1}u^{-1})b(q u/v)}{b(u/v)b(q u v)}, \\
&& y(u,v)= -\frac{1}{b(q v^2)b(quv)}, \quad r(u,v) =\frac{\phi(q^{-1}u^{-1})}{b(v/u)},
\quad q(u,v)=\frac{b(u v)}{b(u/v)b(q u v)}.
\een

Direct calculation gives the following relations, 
\ben\label{idUWT1}
g(u,v)\phi(u)k^\pm(u)+n(u,v)k^\pm(q^{-1}u^{-1})&=&F(u,v)\phi(q^{-1} v^{-1})\phi(v)k^\pm(v),\\
\label{idUWT2}
k(u,v)k^\pm(q^{-1}u^{-1})+w(u,v)\phi(u)k^\pm(u)&=&-F(u,v)\phi(v)k^\pm(q^{-1}v^{-1}).
\een

From the reflection algebra (\ref{RE}), one can extract  the commutations relations between the operators $\mathscr{A}$, $\mathscr{D}$, $\mathscr{C}$ and $\mathscr{B}$.
To order monomials of such operators in the basis span by operator valued series 
\ben
\mathscr{M}_{bdac}(\bar u,\bar v,\bar w,\bar x)=\mathscr{B}(\bar u)\mathscr{D}(\bar v)\mathscr{A}(\bar w)\mathscr{C}(\bar x)
\een
one need the following commutation relations
\begin{eqnarray}
&&\mathscr{A}(u)\mathscr{B}(v)= f(u,v)\mathscr{B}(v)\mathscr{A}(u) + g(u,v)\mathscr{B}(u)\mathscr{A}(v) + w(u,v)\mathscr{B}(u)\mathscr{D}(v),\label{comAB}  \\
&&\mathscr{C}(v)\mathscr{A}(u)=f(u,v) \mathscr{A}(u)\mathscr{C}(v) +g(u,v)\mathscr{A}(v)\mathscr{C}(u) + w(u,v)\mathscr{D}(v)\mathscr{C}(u),\label{comCA}  \\
&&\mathscr{D}(u)\mathscr{B}(v)= h(u,v)\mathscr{B}(v) \mathscr{D}(u) +k(u,v)\mathscr{B}(u)\mathscr{D}(v)+ n(u,v)\mathscr{B}(u)\mathscr{A}(v),\label{comDB} \\
&&\mathscr{C}(v)\mathscr{D}(u)= h(u,v)\mathscr{D}(u) \mathscr{C}(v) +k(u,v)\mathscr{D}(v)\mathscr{C}(u)+n(u,v) \mathscr{A}(v)\mathscr{C}(u),\label{comCD}\\
&&\mathscr{C}(u)\mathscr{B}(v)=\mathscr{B}(v)\mathscr{C}(u)+s(u,v)\mathscr{A}(u)\mathscr{A}(v)+x(u,v)\mathscr{A}(v)\mathscr{A}(u)+y(u,v)\mathscr{D}(u)\mathscr{A}(v)\nonumber\\
&&\qquad\qquad\qquad\quad\quad+r(u,v)\mathscr{A}(u)\mathscr{D}(v)+q(u,v)\mathscr{A}(v)\mathscr{D}(u)+w(u,v)\mathscr{D}(u)\mathscr{D}(v),\label{comCB} \\
&&\mathscr{A}(u)\mathscr{D}(v)=\mathscr{D}(v)\mathscr{A}(u)+k(v,u)\big(\mathscr{B}(u)\mathscr{C}(v)-\mathscr{B}(v)\mathscr{C}(u)\big)\label{comAD}
\end{eqnarray}
and
\ben
&&\mathscr{A}(u)\mathscr{A}(v)=\mathscr{A}(v)\mathscr{A}(u)+w(u,v)\big(\mathscr{B}(u)\mathscr{C}(v)-\mathscr{B}(v)\mathscr{C}(u)\big)
,\label{comAA} \\
&&\mathscr{D}(u)\mathscr{D}(v)=\mathscr{D}(v)\mathscr{D}(u)-\phi(u)\phi(v)w(u,v)\big(\mathscr{B}(u)\mathscr{C}(v)-\mathscr{B}(v)\mathscr{C}(u)\big)
,\label{comDD} \\
&&\mathscr{B}(u)\mathscr{B}(v) = \mathscr{B}(v)\mathscr{B}(u),\label{comBB} \\
&&\mathscr{C}(u)\mathscr{C}(v) = \mathscr{C}(v)\mathscr{C}(u).\label{comCC} 
\end{eqnarray}
Let us remark that this set of relations is complete, {\it i.e.} they are isomorphic to the reflection equation.
We can also define another ordering,
\ben
\mathscr{\hat M}_{cadb}(\bar u,\bar v,\bar w,\bar x)=\mathscr{C}(\bar u)\mathscr{\hat A}(\bar v)\mathscr{\hat D}(\bar w)\mathscr{B}(\bar x).
\een
Here we give only the two most relevant relations for the ABA,
\ben
&&\mathscr{\hat A}(u)\mathscr{C}(v)= h(u,v)\mathscr{C}(v)\mathscr{\hat A}(u) + k(u,v)\mathscr{C}(u)\mathscr{\hat A}(v) + n(u,v)\mathscr{C}(u)\mathscr{\hat D}(v),\label{comAtC}  \\
&&\mathscr{\hat D}(u)\mathscr{C}(v)= f(u,v)\mathscr{C}(v) \mathscr{\hat D}(u) +g(u,v)\mathscr{C}(u)\mathscr{\hat D}(v)+ w(u,v)\mathscr{C}(u)\mathscr{\hat A}(v).\label{comDtC} 
\een 

The dynamical operators (\ref{DyntoK}) can be expressed in terms of the initial operators  (\ref{K}),
\bea\label{DyntoK}
&&\mathscr{B}(u,m)=
\frac{qu}{\gamma_{m+1}}
\Big(\mathscr{B}(u)+
 q^m\beta\big(q u \phi(q^{-1}u^{-1})\mathscr{A}(u)-u^{-1}\mathscr{D}(u)\big)
- (q^{m}\beta)^2\,\mathscr{C}(u)\Big)\\
&&\mathscr{A}(u,m)=\frac{qu}{\gamma_{m-1}}
\Big(\mathscr{B}(u)+q^{m-2}\beta\Big(\big(uq^{2-2m}\frac{\alpha}{\beta} -\frac{u^{-1}}{b(q u^2)}\big )
\mathscr{A}(u)-u^{-1}\mathscr{D}(u)\Big)-q^{-2}\alpha\beta \mathscr{C}(u)\Big)\\
&&\mathscr{D}(u,m)=
\frac{qu}{\gamma_{m-1}}
\Big(q^{m-1}\beta
\Big(
(q^{-2m}u^{-1}
\frac{ \alpha}{\beta}+\frac{u }{b(q u^2)})\mathscr{D}(u)-u\phi(u)\phi(q^{-1}u^{-1})\mathscr{A}(u)
\Big)\\&& \nonumber \qquad\qquad\qquad\qquad\qquad -
\phi(u)(\mathscr{B}(u)-q^{-2}\alpha\beta \mathscr{C}(u))\Big)\\
&&\mathscr{C}(u,m) =\frac{qu}{\gamma_{m-1}}
\Big(q^{-2m}\alpha^2 \mathscr{C}(u)-q^{-m}\alpha\big(q u \phi(q^{-1}u^{-1})\mathscr{A}(u)- u^{-1}\mathscr{D}(u)\big)- \mathscr{B}(u)\Big)
\eea
It allows to find the dynamical commutation relations,
\begin{eqnarray}
&&\mathscr{B}(u,m+2)\mathscr{B}(v,m) = \mathscr{B}(v,m+2)\mathscr{B}(u,m),\label{comBdBd} \\
&&\mathscr{A}(u,m+2)\mathscr{B}(v,m)=f(u,v)\mathscr{B}(v,m)\mathscr{A}(u,m) +
\\\nonumber&&\qquad \qquad g(u,v,m)\mathscr{B}(u,m)\mathscr{A}(v,m) + w(u,v,m)\mathscr{B}(u,m)\mathscr{D}(v,m),\label{comAdBd}  \\
&& \mathscr{D}(u,m+2)\mathscr{B}(v,m)= h(u,v)\mathscr{B}(v,m) \mathscr{D}(u,m) 
\\\nonumber&&\qquad \qquad+k(u,v,m)\mathscr{B}(u,m)\mathscr{D}(v,m)+ n(u,v,m)\mathscr{B}(u,m)\mathscr{A}(v,m),\label{comDdBd}\\
\nonumber
\\
&&\mathscr{C}(u,m-2)\mathscr{C}(v,m) = \mathscr{C}(v,m-2)\mathscr{C}(u,m),\label{comCdCd} \\
&&\mathscr{\hat A}(u,m-2)\mathscr{C}(v,m)= h(u,v)\mathscr{C}(v,m)\mathscr{\hat A}(u,m) 
\\\nonumber&&\qquad \qquad+ \hat k(u,v,m)\mathscr{C}(u,m)\mathscr{\hat A}(v,m) +\hat  n(u,v,m)\mathscr{C}(u,m)\mathscr{\hat  D}(v,m),\label{comhAdCd}  \\
&& \mathscr{\hat D}(u,m-2)\mathscr{C}(v,m)= f(u,v)\mathscr{C}(v,m) \mathscr{\hat D}(u,m) 
\\\nonumber&&\qquad \qquad+\hat g(u,v,m)\mathscr{C}(u,m)\mathscr{\hat D}(v,m)+ \hat w(u,v,m)\mathscr{C}(u,m)\mathscr{\hat A}(v,m).\label{comhDdCd}
\end{eqnarray}
The first term of the commutation relations (\ref{comAdBd},\ref{comDdBd})
and  (\ref{comhAdCd},\ref{comhDdCd}) corresponds to the wanted terms for the ABA.
They have the same form of the
commutation relations for the diagonal case (\ref{comAB},\ref{comDB}) and  (\ref{comAtC},\ref{comDtC}). For the unwanted term we have a new form that depends of $m$ and of the gauge parameters $\alpha$ and $\beta$, namely  
\ben
&&g(u,v,m)=\frac{\gamma(u/v,m+1)}{\gamma_{m+1}}g(u,v), \quad w(u,v,m)=\frac{\gamma(uv,m)}{\gamma_{m+1}}w(u,v),\\
 &&k(u,v,m)=\frac{ \gamma(v/u,m+1)}{\gamma_{m+1}}k(u,v), \quad
n(u,v,m)=\frac{\gamma(1/(uv),m+2)}{\gamma_{m+1}} n(u,v),\\
&&\hat g(u,v,m)=\frac{\gamma(v/u,m-1)}{\gamma_{m-1}}g(u,v), \quad \hat w(u,v,m)=\frac{\gamma(1/(uv),m)}{\gamma_{m-1}}w(u,v),\\
 &&\hat k(u,v,m)=\frac{ \gamma(u/v,m-1)}{\gamma_{m-1}}k(u,v), \quad
\hat n(u,v,m)= \frac{\gamma(uv,m-2)}{\gamma_{m-1}} n(u,v).
\een

\section{Representation theory for triangular boundary case\label{rep}}
For finite dimensional representation of the
quantum one-row monodromy matrix,
we always have a highest weight representation \cite{Tar85}.
For the fundamental representation we used here, the
quantum one-row monodromy matrix is given by the
product $R_{a1}(u/v_1)\dots R_{aN}(u/v_N)$, or by its reflected inverse, and the highest weight vector is
\ben\label{Om}
|\Omega\rangle=\otimes_{k=1}^{N}\left(\begin{array}{l}
       1 \\
      0       \end{array}\right)_k.
     \een
When the right boundary is upper triangular, {\it i.e.},  $\tilde \tau^2=0$,
the vector (\ref{Om}) turns out to be a reference state also for the double-row operators
(\ref{K}). This can be easily seen by expressing the operators 
$\{\mathscr{A}(u),\mathscr{B}(u),\mathscr{C}(u),\mathscr{D}(u)\}$
in terms of the entries of the quantum one-row monodromy matrix,
see {\it e.g.} \cite{Bel14} for explicit formulas.
Then, from their actions on the highest weight vector (\ref{Om})
we find,
\begin{eqnarray}\label{Avac}
 &&\mathscr{A}(u)|\Omega\rangle=k^-(u) \Lambda(u)|\Omega\rangle,\\
&&  \mathscr{D}(u)|\Omega\rangle=\phi(q^{-1}u^{-1})k^-(q^{-1}u^{-1})\Lambda(q^{-1}u^{-1})|\Omega\rangle, \\
&& \mathscr{C}(u)|\Omega\rangle=0.
\end{eqnarray}

\end{document}